\documentclass[prl, twocolumn, superscriptaddress]{revtex4-2}
\usepackage{amsmath,amssymb,bm}
\usepackage{hyperref}
\usepackage{graphicx}
\usepackage{epstopdf}
\usepackage{latexsym}
\usepackage{appendix}
\usepackage{braket}
\usepackage{float}
\usepackage[normalem]{ulem}
\usepackage{comment}
\usepackage{mathtools}
\usepackage{cleveref}
\usepackage{xcolor}

\definecolor{Jerome}{rgb}{0.0, 0.7, 0.3}

\definecolor{Flavio}{rgb}{0.7, 0.0, 0.3}

\usepackage[normalem]{ulem}
\def\BibTeX{{\rm B\kern-.05em{\sc i\kern-.025em b}\kern-.08em
    T\kern-.1667em\lower.7ex\hbox{E}\kern-.125emX}} 
\begin{document}
\title{Effect of inter-edge interaction in a quantum Hall collider}

\author{Amulya Ratnakar}
\affiliation{Aix Marseille Univ, Université de Toulon, CNRS, CPT, Marseille, France}
\author{Flavio Ronetti}
\affiliation{Aix Marseille Univ, Université de Toulon, CNRS, CPT, Marseille, France}
\author{Benoît Grémaud}
\affiliation{Aix Marseille Univ, Université de Toulon, CNRS, CPT, Marseille, France}
\author{Laurent Raymond}
\affiliation{Aix Marseille Univ, Université de Toulon, CNRS, CPT, Marseille, France}
\author{Thierry Martin}
\affiliation{Aix Marseille Univ, Université de Toulon, CNRS, CPT, Marseille, France}
\author{Thibaut Jonckheere}
\affiliation{Aix Marseille Univ, Université de Toulon, CNRS, CPT, Marseille, France}
\author{Jérôme Rech}
\affiliation{Aix Marseille Univ, Université de Toulon, CNRS, CPT, Marseille, France}

\begin{abstract}
Fractional quantum Hall (FQH) colliders measure anyon exchange phases via time-domain braiding, but the $\nu=2/5$ state exhibits an intriguing negative Fano factor, challenging theoretical predictions. Here, we study the effect of inter-edge interactions in a multi-mode FQH collider. We demonstrate that the resulting fractionalization into eigenmodes causes the anyon beam to decompose into correlated and uncorrelated components, which have very distinct behavior in terms of time-domain braiding. We show that the uncorrelated part dominates in the long-junction limit, reversing the tunneling current sign and reproducing the observed negative Fano factor at $\nu=2/5$. Our results highlight the role of interactions and provide a robust interpretation of anyonic braiding in multi-mode systems.
\end{abstract}
\maketitle
\textit{Introduction:} Anyons are quasiparticles (QPs) that interpolate between bosons and fermions, exhibiting
fractional charge and fractional exchange statistics, and emerge as topological excitations in two-dimensional FQH systems~\cite{anyon_leinaas1977theory,anyon_tsui1982two,anyon_arovas1984fractional,anyons_wilczek1982magnetic,anyons_wilczek1982quantum,anyon_halperin1984statistics}. Due to its insulating nature, experimental probing of the FQH bulk presents significant challenges. The edge excitations, however, can inherit bulk topological signatures—such as fractional charge and braiding phase—either through bulk–boundary correspondence or by enabling inter-edge tunneling at a quantum point contact (QPC). A QPC connecting spatially separated quantum Hall edges forces charge transfer to proceed through the intervening bulk, effectively endowing the tunneled edge excitations with bulk topological character. Although fractional charge was experimentally validated decades ago, from Fano factor~\cite{FC_kane1994nonequilibrium,Chamon95,FC_de1998direct,FC_saminadayar1997observation} or photo-assisted transport~\cite{Crepieux04,Intf_chevallier2010photo,Kapfer19,bisognin19}, direct evidence of fractional statistics has continued to be elusive. Recently, the fractional statistics of anyons were experimentally quantified in interferometer~\cite{Intf_campagnano2012hanbury,Intf_chamon1997two,Intf_halperin2011theory,Intf_kim2005signatures,Intf_law2006electronic,Intf_safi2001partition,Intf_guyon2002klein,Intf_guyon2002comptes,Intf_vishveshwara2003revisiting,Intf_kundu2023anyonic,Ronetti25,Ronetti25b,kivelson25} and in collider based~\cite{Collider_glidic2023cross,Collider_lee2023partitioning,Collider_bartolomei2020fractional,Collider_ruelle2023comparing, Collider_ruelle2025time} configurations for various filling fractions.

The geometries based on interferometers, like Fabry-Perot interferometer for a quantum Hall state~\cite{FP_nakamura2020direct,FP_nakamura2023fabry}, involve the spatial braiding of bulk anyons with edge anyons, leading to an interference pattern dependent upon the exchange phase of the involved anyons. In contrast, a FQH collider geometry involves braiding of edge anyons in the time domain~\cite{Collider_lee2023partitioning,TH_colliderrosenow2016current,TH_collider_jonckheere2023anyonic,TH_collider_morel2022fractionalization,TH_Collider_lee2019negative,TH_Collideriyer2024finite,TH_Collider_mora2022anyonic}. The mechanism here involves the convolution of two time-ordered processes, wherein an injected anyon at the source QPC arrives at the collider QPC either before or after the creation of an anyon quasiparticle-quasihole (QP-QH) pair, forming a braiding loop in the time domain.

Experimental measurements have validated theoretical results for Laughlin FQH states, such as $\nu=1/3$ filling~\cite{Collider_bartolomei2020fractional,Collider_glidic2023cross,Collider_nakamura2020direct,Collider_lee2023partitioning}. In contrast, for hierarchical FQH phases supporting multiple edge channels, e.g. at $\nu=2/5$, the theoretical prediction for the exchange phase differs drastically from the experimental results~\cite{Collider_glidic2023cross,Collider_ruelle2023comparing}.  
Although theoretical analysis of the $\nu=2/5$ FQH state predicts a braiding phase of $3\pi/5$ accompanied by a negative tunneling current, several experimental studies report a negative generalized Fano-factor~\cite{Collider_glidic2023cross,Collider_ruelle2023comparing}, indicating a positive tunneling current, thereby highlighting a clear discrepancy between theory and experiments. It was proposed, in previous works~\cite{TH_Collideriyer2024finite,thamm24_finitewidth}, that accounting for the finite width of anyons in the colliding streams could qualitatively explain the experimental results, yet alternative scenarios remain to be explored.

In this Letter, we study the effect of short-range inter-edge interaction on time-domain braiding in a collider geometry. We consider a $\nu=2/5$ FQH collider, which at the edge admits two co-propagating edge modes. We demonstrate that, in the presence of inter-edge interaction, each dilute anyon beam produced through Poissonian injection processes at the source QPCs decomposes into correlated and uncorrelated components while braiding in the time domain with the anyon QP-QH pair at the collider QPC. The correlated part of the anyon beam leads to the usual braiding phase that remains invariant in the presence of interaction, whereas the uncorrelated part of the anyon beam leads to an interaction-dependent braiding phase. We show that in the long junction limit, the contribution from the uncorrelated part of the anyon stream dominates. The resulting effective braiding phase leads to the change in sign of the tunneling current, which in turn results in a negative generalized Fano factor for $\nu=2/5$ FQH collider, compatible with experiments~\cite{Collider_glidic2023cross,Collider_ruelle2023comparing}.

\begin{figure}[ht]
\includegraphics[width=0.48\textwidth]{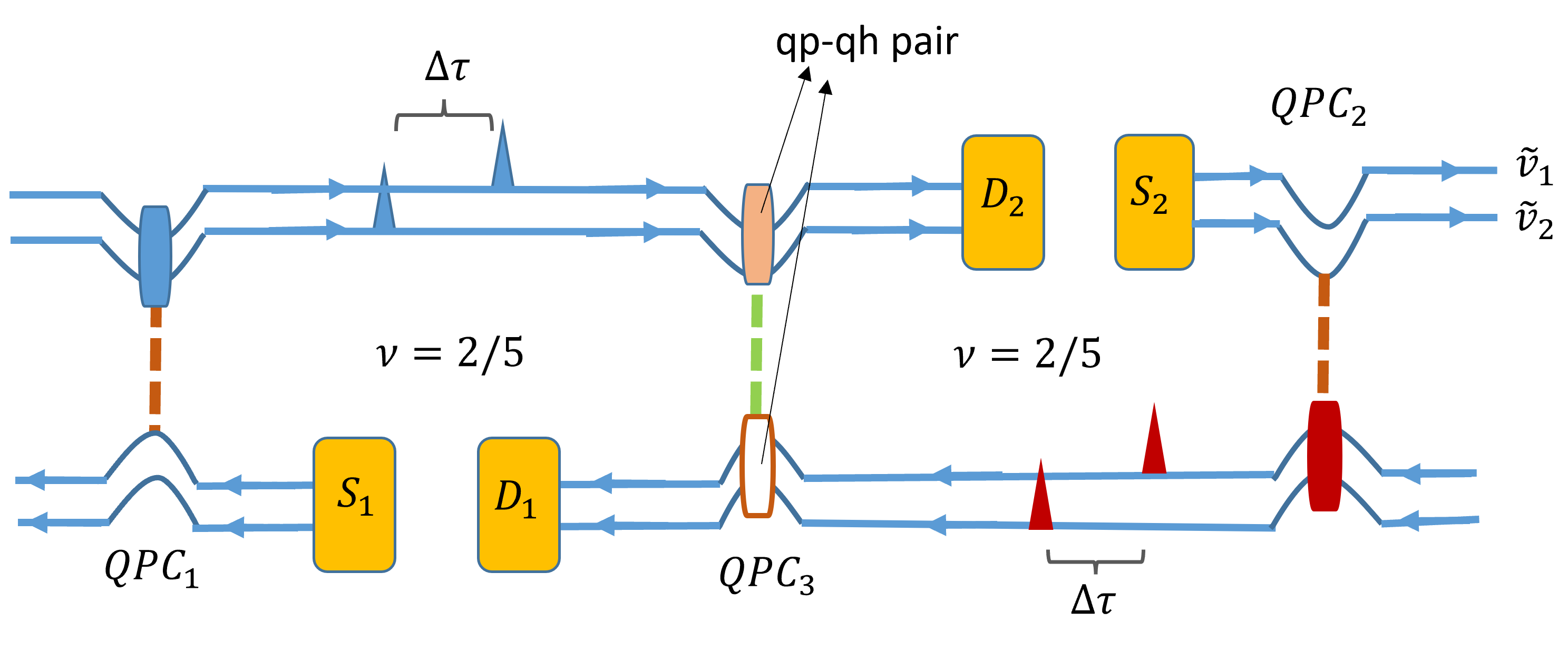}
    \caption{Schematic diagram of the FQH collider geometry at $\nu=2/5$, with two source QPCs, $QPC_{1}$ and $QPC_{2}$ biased by a DC potential at $S_{1}$ and $S_{2}$ respectively. $D_{1}$ and $D_{2}$ are the detectors at which the tunneling current and tunneling noise are measured.}
    \label{fig:setup}
\end{figure}

\textit{Model:} A $\nu=2/5$ FQH state supports two co-propagating edge modes at the edge, with a conductance of $1/3(e^{2}/h)$ and $1/15(e^{2}/h)$ for the outer and inner modes respectively, allowing us to define the electromagnetic coupling vector $\mathbf{Q}=\left(1/\sqrt{3}, 1/\sqrt{15} \right)^T$. Using the bosonization technique~\cite{Bosonization_Haldane_1981,Bosonization_Srao2002,Bosonization_Tgiamarchi2004,Bosonization_von1998,Bosonization_wen1990chiral,Bosonization_wen1990PRL}, the low-energy excitations at the edge can be described in terms of bosonic fields as, $\psi_{\alpha} = (1/\sqrt{2\pi a})e^{i\bm{g}_{\alpha}. \bm{\phi}(x,t)}$~\cite{Bosonization_shtanko2014nonequilibrium}, with $a$ a short distance cutoff. The vector $\bm{\phi}(x,t) = \left( \phi_{1},\phi_{2} \right)^{T}_{(x,t)}$, where, $\phi_{i}(x,t)$ is the bosonic mode corresponding to the $i^{th}$ edge mode, satisfies the standard commutation relation $[\phi_{i}(x),\phi_{j}(y)] = i\pi\delta_{ij}\mathrm{Sgn}(x-y)$. The most relevant quasi-particle operators at the edge are given by $\bm{g}_{A} = \left( 0,\sqrt{3/5} \right)^{T}$ and $\bm{g}_{B} = \left( 1/\sqrt{3},-2/\sqrt{15} \right)^{T}$, defining two types of anyons (labeled $A$ and $B$) with the same charge $e^* = e/5$, exchange phase $\theta = 3\pi/5$, and scaling dimension $\delta = 3/5$. Such a scaling dimension, $\delta > 1/2$, requires to account for the width of the colliding anyonic wavepackets~\cite{TH_Collideriyer2024finite,thamm24_finitewidth}. 
In order to highlight the sole effect of inter-edge interactions, we start by considering extremely thin wave packets, corresponding to very small transmission of the source QPCs. In a second step, we will show the combined effect of interactions and the finite width of colliding anyons. The amount of short-range inter-edge interaction between edge modes is parametrized by $\tilde{u}$; as such, the edge Hamiltonian, in terms of bosonic fields, is given by
\begin{align}
H = \frac{\hbar}{4\pi}\int dx & \left\{ v_{1} \left[\partial_{x}\phi_{1}(x,t) \right]^{2} + v_{2} \left[ \partial_{x}\phi_{2}(x,t) \right]^{2} \right. \nonumber\\
& \left. + 2 \tilde{u} \partial_{x}\phi_{1}(x,t)\partial_{x}\phi_{2}(x,t) \right\},
\label{eq:generalH}
\end{align}
where $v_{i}$ is the bare velocity of the $i^{th}$ edge mode.

The Hamiltonian can be diagonalized in terms of bosonic eigenmodes $\tilde{\phi}_{i}(x,t)$, which are related to the interacting bosonic modes $\phi_{i}(x,t)$ by a diagonalizing matrix $M$~\cite{Int_agarwal2009enhancement,Int_Amulya,Int_das2009effect,Int_ratnakar2021enhancement}, such that $\bm{\phi} (x,t) = M \tilde{\bm{\phi}} (x,t)$. This basis rotation preserves the commutation relations of the bosonic fields and transforms the vectors $\mathbf{Q}^T$ and $\bm{g}^T$ into $\tilde{\mathbf{Q}}^T=\mathbf{Q}^T M$ and $\tilde{\bm{g}}^T= \bm{g}^T M$. The renormalized velocities of the eigenmodes are given by
\begin{equation}
\tilde{v}_{1/2} = \frac{1}{2}\left( v_{1} + v_{2} \pm \sqrt{(v_{1}-v_{2})^{2} + 4\tilde{u}^{2}} \right).
\end{equation}
Since $\tilde{v}_1>\tilde{v}_2$, the fields $\tilde{\phi}_1$ and $\tilde{\phi}_2$ describes the fast and slow eigenmode respectively. In what follows, we make the simplifying assumption $v_1 = v_2 = v$.

\textit{Single QP injection:} The collider geometry consists of a FQH state at filling $\nu=2/5$, whose counter-propagating edge states at the upper and (lower) boundary are labeled with the superscript $U (L)$. The two source QPCs ($QPC_{1/2}$) are located at $x = \pm L$ from the collider QPC ($QPC_{3}$), which sits at $x = 0$ (see Fig.~\ref{fig:setup}c). All three QPCs are in the weak tunneling regime, which allows the tunneling of the most relevant quasi-particle ($A$ and $B$-type anyons) through the bulk. The source QPCs, $QPC_{1}$ and $QPC_{2}$, are biased such that they inject anyons into the upper and lower edges, respectively. 

Before addressing the case of a dilute anyon stream, we first consider the prepared state $\left| \varphi_{\gamma,\xi} \right\rangle = {\psi_\gamma^U}^\dagger \left(-L,-T^U \right) {\psi_\xi^L}^\dagger \left(L,-T^L\right) \ket{0}$, which replaces the ground state when calculating all the expectation values of the observables~\cite{TH_collider_jonckheere2023anyonic,multikeldysh}. Here, ${\psi_\alpha^{U(L)}}^\dagger \left(x,-T^{U(L)}\right)$ is the quasiparticle creation operator, creating an $\alpha$-type anyon at position $x$ and time $-T^{U(L)}$ along the upper (lower) edge. In this way, the effect of interactions appears more clearly; later, we will introduce the Poissonian stream.

The tunneling Hamiltonian, $H_{T}$, and tunneling current operator, $\hat{I}_{T}$, at $QPC_{3}$, are given by $H_{T} = \sum_{\alpha} \Gamma_{\alpha} \left( {\psi_\alpha^U}^\dagger(0,t)\psi_\alpha^L (0,t) + \text{H.c.} \right)$ and $\hat{I}_{T} =i e^* \sum_{\alpha} \Gamma_{\alpha}\left( {\psi_\alpha^U}^\dagger(0,t)\psi_\alpha^L (0,t) - \text{H.c.} \right)$, respectively. Using the Keldysh technique, the expectation value of the tunneling current can be calculated to the lowest order in the coupling $\Gamma_{\alpha}$, such that
\begin{align}
\langle\hat{I}(t)\rangle =& -4 e^* \sum_{\alpha} |\Gamma_{\alpha}|^{2} \sum_{\gamma, \xi}  \int_0^{\infty} d\tau ~\text{Im}\left[ \tilde{G}(\tau)^{2 \delta} \right] \nonumber\\
& \times \sin \left[ 2\pi \Lambda_{\alpha \gamma} \left(t,\tau,T^{U}\right)-2\pi\Lambda_{\alpha \xi} \left(t,\tau,T^{L} \right) \right],
\label{Eq:IT_ST}
\end{align}   
where the scaling dimension is not affected by the basis change, $\delta = \bm{g}_\alpha.\bm{g}_\alpha = \tilde{\bm{g}}_\alpha.\tilde{\bm{g}}_\alpha$. The expression for the current involves the time-dependent braiding phases
\begin{align}
\Lambda_{\alpha \gamma} \left( t,\tau,T^{U/L} \right) = & \sum_{j=1}^{2} \tilde{g}_{\alpha,j}\tilde{g}_{\gamma,j} \left[\Theta \left(T^{U/L} - \frac{L}{\tilde{v}_j} + t \right) \right. \nonumber \\
& \left. -\Theta \left(T^{U/L}- \frac{L}{\tilde{v}_j} + t-\tau \right)\right],
\label{Eq:Lambda}
\end{align} 
which results from the braiding of the injected $\gamma$-type anyon with the $\alpha$-type QP-QH pair created at $QPC_{3}$, over the time window $-T^{U/L}+L/\tilde{v}_{j} < t < -T^{U/L}+L/\tilde{v}_{j} + \tau$. Incidentally, because the eigenmodes propagate with different velocities, this cannot be written only in terms of the mutual braiding phase of $\alpha$ and $\gamma$-type anyons but instead involves individual components over each eigenmode. $\tilde{G}(\tau)$ is the standard bosonic Green's function for the free eigenmodes
\begin{equation}
\tilde{G}(\tau) = \frac{\sinh \left(i\pi\tau_0/(\hbar\beta)\right)}{\sinh \left[ (\pi/\hbar\beta)(i\tau_0-\tau) \right]},
\end{equation} 
with $\beta$ the inverse temperature and $\tau_0$ a short time cutoff.

The tunneling current results from time-domain braiding, an interference effect between processes where QP-QH pair tunneling at the QPC occurs before or after the arrival of the incoming anyon~\cite{Collider_lee2023partitioning,TH_colliderrosenow2016current,TH_collider_jonckheere2023anyonic,TH_collider_morel2022fractionalization,TH_Collider_lee2019negative,TH_Collideriyer2024finite,TH_Collider_mora2022anyonic}.
From Eq.~\eqref{Eq:IT_ST}, one readily sees that the tunneling current at time $t$ has a contribution from every earlier time $t'$. It involves the anyon propagator and the sine of the braiding phase from all the anyons having reached the QPC between times $t'$ and $t$. As a consequence of interactions, an injected anyon fractionalizes into two parts (corresponding to the two eigenmodes) propagating at different velocities $\tilde{v}_{1,2}$, reaching the QPC at different times, and thus contributing partially to the braiding phase.

\begin{figure}
\centering
\includegraphics[scale=0.23]{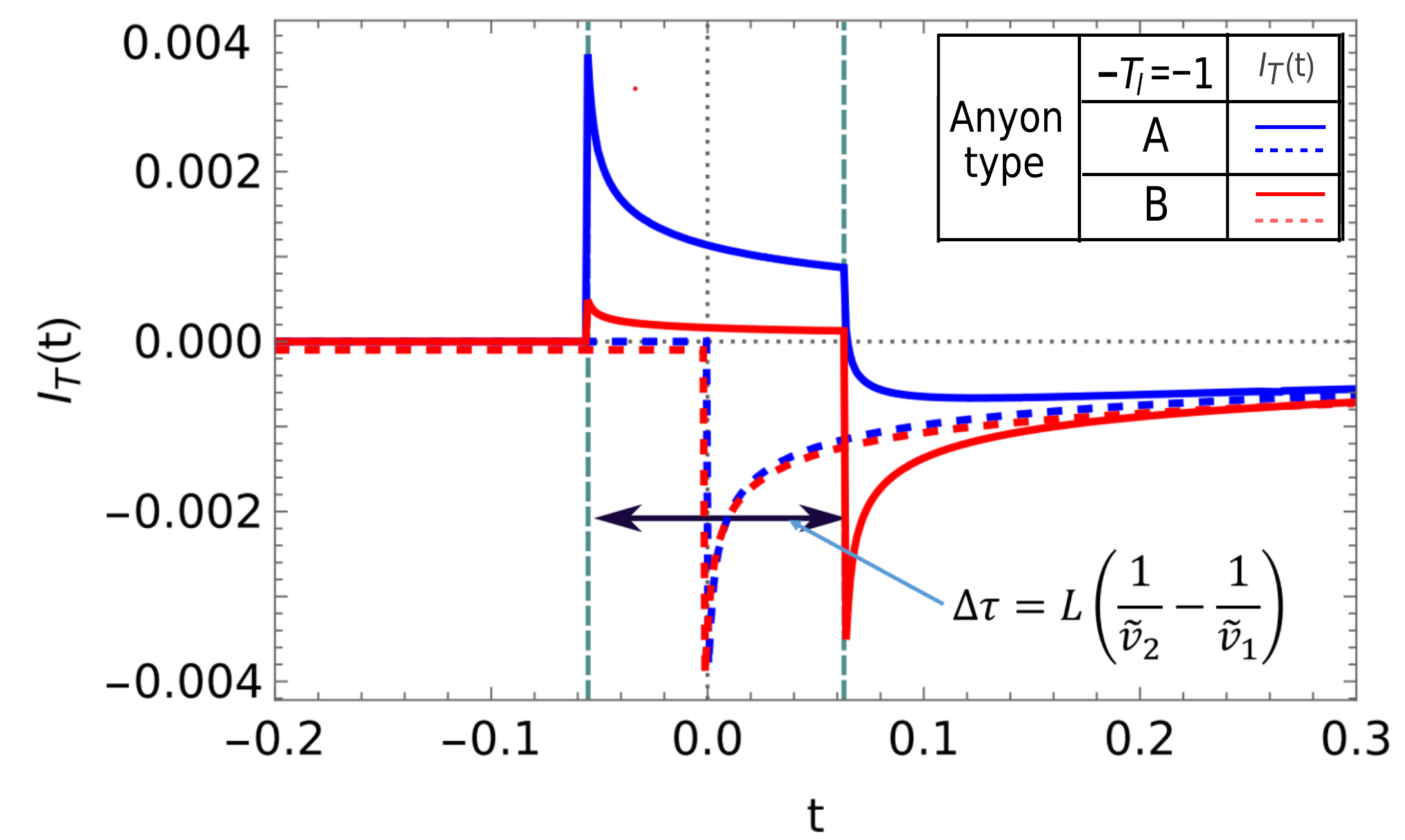}
    \caption{Time-dependent tunneling current for a controlled injection of $A$ and $B$-type anyons only in the upper edge for bare velocities $v_{1}=v_{2} =v$. The blue (red) lines denote the tunneling current when an $A$-type ($B$-type) anyon is injected at $-T^{U} = -1$, with inverse temperature $\hbar \beta=10$, and $\tau_0 = 0.001$ (all times being expressed in units of $L/v$). $u = 0$ and $0.4$ cases are denoted by dashed and solid lines. Dashed green vertical lines denote the time $-T^U+L/\tilde{v}_{1/2}$ at which anyons along the eigenmodes hit $QPC_{3}$.}
    \label{fig:Time-dep_TC}
\end{figure}

This is illustrated in Fig.~\ref{fig:Time-dep_TC},
which shows the time-dependent tunneling current in the low temperature regime ($\hbar \beta > L/v$), when a single anyon is injected, either of type $A$ (blue curves) or $B$ (red curves). In the absence of interactions (dashed curves), the anyon reaches the QPC at time $t=0$. The resulting tunneling current is the same for both anyon types as they share the same properties (exchange phase and scaling dimension), jumping from zero before the arrival of the anyon, to a value proportional to $\sin \left( 2 \pi \tfrac{3}{5} \right)<0$ at $t=0$, before vanishing very slowly.
 
When interaction is switched on, the injected anyon fractionalizes onto the two eigenmodes, with different velocities. When reaching the central QPC, the part along the fast eigenmode contributes only a fraction of the total braiding phase, leading to a positive current (with different values for $A$ and $B$-type anyon) which then slowly decreases. After a time $\Delta\tau = L \left( \frac{1}{\tilde{v}_2} - \frac{1}{\tilde{v}_1} \right)$, the slow mode arrives at the QPC, adding its fraction of the total braiding phase, leading to a negative current, which then slowly decreases in absolute value.

\textit{Poissonian averaging:} 
We now turn to the case of incoming anyons emitted by the two upstream source QPCs operating in the weak tunneling regime. Anyon tunneling events are thus independent and stochastic. 
The tunneling current at the central QPC is then obtained by
generalizing the result in Eq.~\eqref{Eq:IT_ST} to the case of multiple incoming anyons and by averaging over their number, which follows a Poisson distribution.

\begin{figure}
\includegraphics[width=0.48\textwidth]{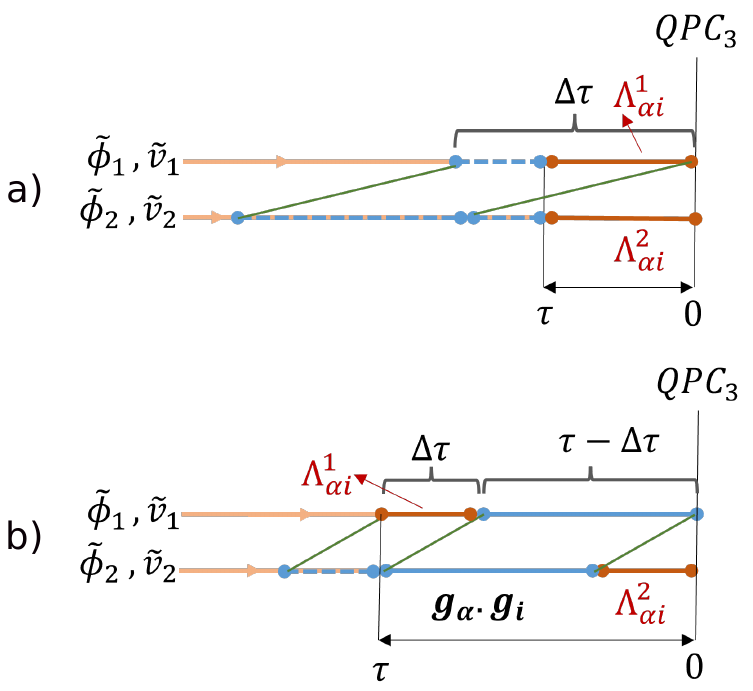}
    \caption{
    Schematic representation of the contributions to the braiding phase from excitations reaching the QPC during the time interval $[0,\tau]$. Because of the different renormalized velocities, propagation along the eignmodes occur with a time delay $\Delta\tau$.
    a) When $\tau < \Delta \tau$, the contributions to the braiding phase correspond to uncorrelated tunneling events at the source, during the entire time period $[0, \tau]$ (shown in red solid lines).
    b) When $\tau > \Delta \tau$, parts of the two streams are correlated, contributing with an invariant phase $\bm{g}_{\alpha}.\bm{g}_{\gamma}$ over the time period $[0,\tau-\Delta\tau]$ (shown in solid blue line).
    The remaining uncorrelated parts of the anyon streams contribute to the braiding phase as $\Lambda^{j}_{\alpha \gamma}$ over a period $\Delta\tau$ (shown in red solid line).
     In both cases, the dashed blue lines show the part of the anyon stream which does not participate in the time-domain braiding over the time period $[0,\tau]$.
    }  \label{fig:short_long_junction_limit}
\end{figure}

Similarly to what has been discussed for the results of Eqs.~\eqref{Eq:IT_ST} and \eqref{Eq:Lambda}, the quantity of interest is the total braiding phase due to the excitations reaching the QPC during the time interval $[0,\tau]$.
The injected dilute stream of anyons is split by the interactions over the two eigenmodes, yielding two trains of anyons separated by a time $\Delta \tau = L(1/\tilde v_2 - 1/\tilde v_1)$.
Depending on the relative values of $\tau$ and $\Delta \tau$, two different regimes emerge. 
For $\Delta \tau >\tau$, the two trains of anyons reaching the QPC during the time interval $[0,\tau]$ are totally uncorrelated, as they have been emitted at different times in stochastically independent tunneling events (see Fig.~\ref{fig:short_long_junction_limit}a). The Poissonian averages are then independent for the two modes. 
On the other hand, for $\tau > \Delta \tau$, there exists a time interval of length $\tau-\Delta \tau$
associated with tunneling events from the source for which both the fast and the slow trains of anyons reach the central QPC during the time interval $[0,\tau]$. For these, a single Poissonian average has to be performed, and we thus refer to them as the correlated part. Outside this time interval, only uncorrelated streams reach the QPC (see Fig.~\ref{fig:short_long_junction_limit}b). 
 Accounting for these two regimes, the Poisson-averaged tunneling current is given by~\footnote{See Supplementary Material for details of the calculation.}
\begin{widetext}
\begin{align}
\langle I_{T} \rangle =& -4 e^* \sum_{\alpha} |\Gamma_{\alpha}|^{2}\int_{0}^{\infty} d\tau ~ \text{Im} \left[ \tilde{G}(\tau)^{2\delta} \right]  
\left[ 
\frac{\Theta(\Delta\tau-\tau)\sin \left\{ \sum_{\gamma}\frac{I_{-}}{e^*}P_{\gamma}\tau \left[\sin(2\pi \Lambda^{1}_{\alpha \gamma}) + \sin(2\pi \Lambda^{2}_{\alpha \gamma}) \right] \right\}}{\exp{\left\{ \sum_{\gamma} \frac{I_{+}}{e^*}P_{\gamma}\tau \left[ 2 - \cos \left(2\pi\Lambda^{1}_{\alpha \gamma} \right) -\cos \left( 2\pi\Lambda^{2}_{\alpha \gamma} \right) \right] \right\}}} 
\right.
\nonumber\\
& + \left. 
 \frac{
\Theta(\tau-\Delta\tau)
\sin \left( \sum_{\gamma}\frac{I_{-}}{e^*}P_{\gamma} \left\{ \Delta\tau \left[\sin(2\pi \Lambda^{1}_{\alpha \gamma}) + \sin(2\pi \Lambda^{2}_{\alpha \gamma})\right] + (\tau-\Delta\tau)\sin(2\pi \bm{g}_{\alpha}.\bm{g}_{\gamma}) \right\} \right)}
{\exp{\left(\sum_{\gamma} \frac{I_{+}}{e^*}P_{\gamma} \left\{ \tau+\Delta\tau - \Delta\tau \left[\cos\left(2\pi\Lambda^{1}_{\alpha \gamma}\right) + \cos \left(2\pi\Lambda^{2}_{\alpha \gamma}\right)\right] -(\tau-\Delta\tau)\cos\left(2\pi\bm{g}_{\alpha}.\bm{g}_{\gamma}\right)  \right\} \right)}}
\right],
\label{Eq:Final_T_current_noise}
\end{align}       
\end{widetext}
where $\Lambda^{j}_{\alpha \gamma}=\tilde{g}_{\alpha,j}\tilde{g}_{\gamma,j}$. Here we introduced $I_\pm = I^U \pm I^L$, with $I^{U/L}$ the average current across $QPC_{1/2}$. The proportion $I_\alpha^{U/L}$ of $\alpha$-type anyon in the injected current is encoded in $P_\alpha$ as $I_{\alpha}^{U/L} = I^{U/L} P_{\alpha}$, with $\sum_{\alpha}P_{\alpha}=1$. The Poisson-averaged zero-frequency tunneling noise $\langle S_T \rangle = \int dt \langle \delta I_T (0) \delta I_T (t) \rangle$, with $\delta I_T (t) = I_T (t) - \langle I_T \rangle$, can be derived similarly~\cite{Note1}.

The value of the tunneling current then results from the competition between two timescales: the temporal separation $\Delta\tau$ between eigenmodes, and the decay time, $T_{\mathrm{decay}}$, of the integrand, which is associated with both the Green's function and the exponential factor. Note that $T_{\mathrm{decay}}$ is a complicated function of the external parameters, which depends on the temperature, the scaling dimension, the effective charge and the total injected current.
In the limiting regime $\Delta\tau \ll T_{\mathrm{decay}}$, the integral in Eq.~\eqref{Eq:Final_T_current_noise} is dominated by the second term, $\tau > \Delta \tau$ (see Fig.~\ref{fig:short_long_junction_limit}b). Consequently, the tunneling current is governed by the invariant exchange phase $\sin(2\pi \bm{g}_\alpha.\bm{g}_\gamma)$ (here $\sin(6\pi/5)<0$) and remains negative. 
In contrast, in the long junction regime (associated with larger values of $L$ at fixed interaction $u$), where $\Delta\tau \gg T_{\mathrm{decay}}$, the first term, $\tau < \Delta \tau$, now dominates the integral in Eq.~\eqref{Eq:Final_T_current_noise} (see Fig.~\ref{fig:short_long_junction_limit}a). The tunneling current is then determined 
 by the interaction-dependent braiding phase $\sum_{j=1}^{2}\sin \left(2\pi\Lambda_{\alpha \gamma}^j\right)$. In the next section, we will show that the latter yields a positive current for a finite interaction strength.

\textit{Generalized Fano factor: }
For simplicity, we now consider only the presence of A-type anyons. This assumption is experimentally well motivated, as these anyons only reside on the inner channel of the edge state and conductance measurements at the QPC suggest that tunneling only involves this channel at low transmission. 

An experimentally accessible quantity proven to be sensitive to the ayonic statistics is the generalized Fano factor~\cite{TH_colliderrosenow2016current}. It is given by the current cross-correlations at the output of the central QPC $\langle \delta I_{D_1}\delta I_{D_2} \rangle$, normalized by the noise in incoming currents transmitted through the same QPC. It reads
\begin{equation}
P \left( \frac{I_-}{I_+} \right) = \frac{\langle\delta I_{D_1}\delta I_{D_2}\rangle}{\left. e^* I_+ \frac{\partial}{\partial I_-} \langle I_{T} \rangle \right|_{I_-=0},
}    
\end{equation}
and can be readily obtained from the computation of the tunneling current $\langle I_T \rangle$ and the tunneling noise $\langle S_T \rangle$~\cite{Note1}.

\begin{figure}
\includegraphics[width=0.48\textwidth]{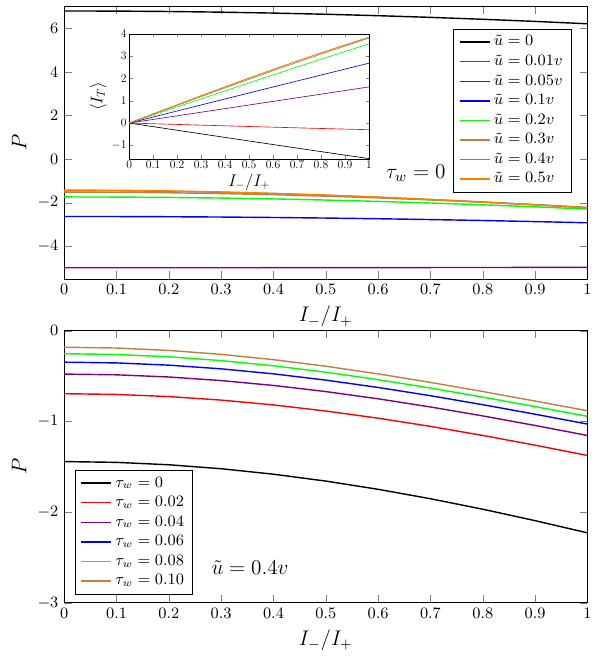}
\caption{Generalized Fano factor $P$, as a function of injection current asymmetry $I_{-}/I_{+}$ for bare velocities $v_{1}=v_{2}=v$, for infinitely thin wavepackets ($\tau_w=0$) and different values of the interaction strength (top), and for $\tilde{u}=0.4v$ and different width of the colliding wavepackets (bottom). The inset shows the tunneling current $\langle I_{T}\rangle$ (in units of $I_{b} = e |\Gamma|^{2}\tau_0^{2\delta} \beta^{1-2\delta}$).}
\label{fig:final_current_noise_P_value}
\end{figure}

Figure~\ref{fig:final_current_noise_P_value} (top) presents the generalized Fano factor $P$ and the tunneling current $\langle I_T\rangle$ (inset) as a function of the injection-current asymmetry $I_-/I_+$ for various values of the inter-edge interaction strength $\tilde{u}$ (similar results for the tunneling noise and cross-correlations can be found in~\cite{Note1}). The calculations are performed using experimentally relevant parameters: $v_1=v_2=v=2\times10^{4}$ m/s, junction length $L=2~\mu$m, temperature $T=30$ mK~\cite{Collider_glidic2023cross,Collider_ruelle2023comparing,exp_val}. 
In the noninteracting limit ($\tilde{u}=0$), the tunneling current is negative, and the Fano factor at symmetric injection ($I_-=0$) is $P\simeq6$, theoretically consistent with a mutual exchange phase of $3\pi/5$~\cite{Note1}. As we increase the interaction strength, the uncorrelated  components  of  the  Poissonian anyon  streams start to dominate the time-domain braiding process. As a result, the tunneling current continuously evolves from negative to positive values. Near the interaction strength where the tunneling current vanishes, the Fano factor diverges (not shown), while for stronger interactions it becomes negative and approaches $P\simeq-1$~\cite{Note1}. 

Since the interpretation of state-of-the-art experiments suggests that anyons have a finite spatial extent rather than being point-like in the collider geometry at $\nu=2/5$, we now reinstate the finite width of the wave packets emitted from the source (see end matter). Adapting the calculation of Refs.~\cite{TH_Collideriyer2024finite} and \cite{thamm24_finitewidth}, we compute the generalized Fano factor in the presence of interactions for Lorentzian wavepackets of increasing width~\cite{Note1}. Results are shown in Fig.~\ref{fig:final_current_noise_P_value} (bottom). The generalized Fano factor for finite width anyons is further shifted toward positive values, but the picture presented above does not qualitatively change, as the sign of the considered quantity stays negative. The experimentally observed values of the generalized Fano factor~\cite{Collider_glidic2023cross,Collider_ruelle2023comparing,exp_val} could thus be attributed to the interplay between finite interaction strength and finite width of the anyon wavepackets emitted from the source.

In summary, we investigated the role of inter-edge interactions on time-domain braiding in a quantum Hall collider geometry. We demonstrated that time-domain braiding involves contributions from both correlated and uncorrelated components of the Poissonian anyon streams propagating along the edge eigenmodes and braiding with QP-QH excitations at the collider QPC. The uncorrelated component gives rise to a braiding phase that is sensitive to inter-edge interactions. For a $\nu=2/5$ QH collider, we showed that the experimentally observed value could emerge from the combined effects of strong inter-edge interactions, finite width of the anyonic wavepackets and tunneling of anyons which reside on the innermost $1/15$ edge channel. 
Finally, our framework is readily extendable to collider configurations with multiple edge modes, including co-propagating and counter-propagating channels, and offers new insights into the interplay between inter-edge interactions and fractional statistics in quantum Hall systems.      

\begin{acknowledgments}
We thank G. F\`eve, G. M\'enard, F. Parmentier and F. Pierre for useful discussions. 
This work was carried out in the framework of the
project ``ANY-HALL'' (Grant No. ANR-21-CE30-0064-03). We acknowledge
funding from the Agence Nationale de la Recherche under
the France 2030 programme, reference ANR-22-PETQ-0012, and from the ERC advanced grant
``ASTEC'' (Grant No. 101096610).
\end{acknowledgments}

\bibliography{citations}

\clearpage

\appendix

\onecolumngrid
\begin{center}
{\bfseries \large End Matter}
\end{center}
\vspace{1ex}
\twocolumngrid

\section{Diagonalizing the interacting Hamiltonian}

Following Refs.~\onlinecite{Bosonization_wen1990chiral} and \onlinecite{Bosonization_wen1990PRL}, the general form of the Hamiltonian, Eq.~\eqref{eq:generalH} can be diagonalized in terms of bosonic eigen-modes $\tilde{\phi}_{i}(x,t)$, which are related to the interacting bosonic modes $\phi_{i} (x,t)$ by a diagonalizing matrix $M$~\cite{Int_agarwal2009enhancement,Int_Amulya,Int_das2009effect,Int_ratnakar2021enhancement}, such that,
\begin{eqnarray}
\begin{pmatrix}
     \phi_1 \\
     \phi_2
 \end{pmatrix}_{(x,t)} &=& M \begin{pmatrix}
       \tilde{\phi}_{1} \\
       \tilde{\phi}_{2}
       \end{pmatrix}_{(x,t)} =  \begin{pmatrix}
      a & b\\
      b & a
      \end{pmatrix}\begin{pmatrix}
       \tilde{\phi}_{1} \\
       \tilde{\phi}_{2}
       \end{pmatrix}_{(x,t)}\nonumber
\end{eqnarray}
where,
\begin{eqnarray}
a &=& \sqrt{\frac{v_{1}-v_{2}+\sqrt{(v_{1}-v_{2})^{2}+4\tilde{u}^{2}}}{2\sqrt{(v_{1}-v_{2})^{2}+4\tilde{u}^{2}}}},\nonumber\\
b &=& \frac{\sqrt{2}\tilde{u}}{\sqrt{\left((v_{1}-v_{2})\sqrt{(v_{1}-v_{2})^{2}+4\tilde{u}^{2}} + (v_{1}-v_{2})^{2}+4\tilde{u}^{2}\right)}},\nonumber\\    
\end{eqnarray}
and the renormalized velocities are given by
\begin{equation}
\tilde{v}_{1/2} = \frac{1}{2}\left( v_{1} + v_{2} \pm \sqrt{(v_{1}-v_{2})^{2} + 4\tilde{u}^{2}} \right).    
\end{equation}
Note that the presence of interactions modifies the vectors $\bm{Q}$ and $\bm{g}_{\alpha}$ to $\bm{\tilde{Q}} = M^{T}\bm{Q}$ and $\bm{\tilde{g}}_{\alpha} = M^{T}\bm{g}_{\alpha}$ along the eigenmodes. However, the interactions do not change the total mutual exchange phase ($\bm{g}_{\alpha}.\bm{g}_{\alpha} = \bm{\tilde{g}}_{\alpha}.\bm{\tilde{g}}_{\beta}$), charge ($q_{\alpha} =\bm{g}_{\alpha}.\bm{Q} = \bm{\tilde{g}}_{\alpha}.\bm{\tilde{Q}}$), and the scaling dimensions of the quasi-particle operator ($\delta_{\alpha} = (1/2)\bm{g}_{\alpha}.\bm{g}_{\alpha} = (1/2)\bm{\tilde{g}}_{\alpha}.\bm{\tilde{g}}_{\alpha}$).

\section{Poisson-averaged tunneling noise}

In the regime where anyons are emitted by the two upstream source QPCs, we obtained the expression for the Poisson-averaged tunneling current , reported in Eq.~\eqref{Eq:Final_T_current_noise}. A similar calculation allows one to derive the expression for the Poisson-averaged tunneling noise, which yields
\begin{widetext}
\begin{align}
\langle S_{T} \rangle =& 4 \left( e^* \right)^2 \sum_{\alpha} |\Gamma_{\alpha}|^{2}\int_{0}^{\infty} d\tau ~ \text{Re} \left[ \tilde{G}(\tau)^{2\delta} \right]  
\left[ 
\frac{\Theta(\Delta\tau-\tau)\cos \left\{ \sum_{\gamma}\frac{I_{-}}{e^*}P_{\gamma}\tau \left[\sin(2\pi \Lambda^{1}_{\alpha \gamma}) + \sin(2\pi \Lambda^{2}_{\alpha \gamma}) \right] \right\}}{\exp{\left\{ \sum_{\gamma} \frac{I_{+}}{e^*}P_{\gamma}\tau \left[ 2 - \cos \left(2\pi\Lambda^{1}_{\alpha \gamma} \right) -\cos \left( 2\pi\Lambda^{2}_{\alpha \gamma} \right) \right] \right\}}} 
\right.
\nonumber\\
& + \left. 
 \frac{
\Theta(\tau-\Delta\tau)
\cos \left( \sum_{\gamma}\frac{I_{-}}{e^*}P_{\gamma} \left\{ \Delta\tau \left[\sin(2\pi \Lambda^{1}_{\alpha \gamma}) + \sin(2\pi \Lambda^{2}_{\alpha \gamma})\right] + (\tau-\Delta\tau)\sin(2\pi \bm{g}_{\alpha}.\bm{g}_{\gamma}) \right\} \right)}
{\exp{\left(\sum_{\gamma} \frac{I_{+}}{e^*}P_{\gamma} \left\{ \tau+\Delta\tau - \Delta\tau \left[\cos\left(2\pi\Lambda^{1}_{\alpha \gamma}\right) + \cos \left(2\pi\Lambda^{2}_{\alpha \gamma}\right)\right] -(\tau-\Delta\tau)\cos\left(2\pi\bm{g}_{\alpha}.\bm{g}_{\gamma}\right)  \right\} \right)}}
\right],
\label{Eq:Final_T_noise}
\end{align}       
\end{widetext}
where $\Lambda^{j}_{\alpha \gamma}=\tilde{g}_{\alpha,j}\tilde{g}_{\gamma,j}$. Here we introduced $I_\pm = I^U \pm I^L$, with $I^{U/L}$ the average current across $QPC_{1/2}$.

\section{Computing the generalized Fano factor}

The generalized Fano factor is an experimentally accessible quantity proven to be sensitive to the ayonic statistics. It is defined as the ratio of the current cross-correlations at the output of the central QPC $\langle \delta I_{D_1}\delta I_{D_2} \rangle$, to the noise in incoming currents transmitted through the same QPC. In the experimentally-motivated situation where we only consider  the presence of $A$-type anyons, it takes the form
\begin{equation}
P \left( \frac{I_-}{I_+} \right) = \frac{\langle\delta I_{D_1}\delta I_{D_2}\rangle}{\left. e^* I_+ \frac{\partial}{\partial I_-} \langle I_{T} \rangle \right|_{I_-=0}}    ,
\end{equation}
where the  current cross-correlations at the output can be written as~\cite{TH_colliderrosenow2016current}
\begin{align}
\langle\delta I_{D_1}\delta I_{D_2}\rangle = - \langle S_{T} \rangle + e^* \left(  I_+  \frac{\partial}{\partial  I_- }  +  I_- \frac{\partial}{\partial  I_+ } \right) \langle I_{T} \rangle .
\end{align}
The tunneling current and the tunneling noise can be readily ontained from Eqs.~\eqref{Eq:Final_T_current_noise} and \eqref{Eq:Final_T_noise} respectively, upon substituting $P_A = 1$ and $P_B = 0$, in order to consider the sole presence of $A$-type anyons.

\section{Saturation of the generalized Fano factor}

Both the tunneling current and the generalized Fano factor saturate once the interaction goes beyond a threshold value.
This can be understood from the competition of timescales between $\Delta \tau$ and $T_\text{decay}$.
As the interaction strength increases, so does the timescale $\Delta \tau$, going from a regime dominated by the correlated components of the anyon streams to one that involves mainly the uncorrelated parts.
Once $\Delta \tau \sim T_\text{decay}$, the current is fully controlled by the first term in Eq.~\eqref{Eq:Final_T_current_noise} as the short time behavior dominate the integral, and increasing the interaction further only has a limited impact on the value of the current. One can work out the leading contribution to the generalized Fano factor in this case~\cite{Note1} yielding, at zero temperature,
\begin{align}
    P(0) = 1 - \frac{\cot \pi\delta}{1 - 2 \delta} \frac{2-\cos \left( 2 \pi \Lambda_{AA}^1\right) - \cos \left( 2 \pi \Lambda_{AA}^2\right)}{\sin \left( 2 \pi \Lambda_{AA}^1\right) + \sin \left( 2 \pi \Lambda_{AA}^2\right)} .
\end{align}
Noticing that, in the equal velocity case $v_1=v_2$, one has $\Lambda_{AA}^1 = \Lambda_{AA}^2 = 3/10$ independently of $u$, finally leads to $P(0) \simeq -1.236$, as obtained in Fig.~\ref{fig:final_current_noise_P_value}b.

\section{Finite width wavepackets}
Here, we account for the finite width of the injected anyons across the source QPCs, closely following previous works~\cite{TH_colliderrosenow2016current,TH_Collideriyer2024finite,thamm24_finitewidth}. 

The tunneling current and the tunneling noise can then be calculated as
\begin{widetext}
\begin{align}
\langle I_{T} \rangle =& -4 e^*  |\Gamma_A|^{2}\int_{0}^{\infty} d\tau ~ \text{Im} \left[ \tilde{G}(\tau)^{2\delta} \right]  
\left[ 
\frac{\Theta(\tau-\Delta\tau)\sin \left( \frac{I_{-}}{e^*} \left\{\text{Im}[F^{1}(\Delta\tau)] + \text{Im}[F^{2}(\Delta\tau)] + \text{Im}[F^{T}(\tau-\Delta\tau)] \right\} \right)}{\exp{\left(  \frac{I_{+}}{e^*} \left\{ \text{Re}[F^{1}(\Delta\tau)] + \text{Re}[F^{2}(\Delta\tau)] + \text{Re}[F^{T}(\tau-\Delta\tau)] \right\} \right)}} 
\right.
\nonumber\\
& + \left. 
 \frac{
\Theta(\Delta\tau - \tau)
\sin \left( \frac{I_{-}}{e^*}\left\{ \text{Im}[F^{1} (\tau)] + \text{Im}[F^{2}(\tau)] \right\} \right)}
{\exp{\left(\frac{I_{+}}{e^*}\left\{ \text{Re}[F^{1}(\tau)] + \text{Re}[F^{2}(\tau)] \right\} \right)}}
\right],
\label{Eq:Final_T_current_noise2} \\
\langle S_{T} \rangle =& 4 \left(e^*\right)^2  |\Gamma_A|^{2}\int_{0}^{\infty} d\tau ~ \text{Re} \left[ \tilde{G}(\tau)^{2\delta} \right]  
\left[ 
\frac{\Theta(\tau-\Delta\tau) \cos \left( \frac{I_{-}}{e^*} \left\{\text{Im}[F^{1}(\Delta\tau)] + \text{Im}[F^{2}(\Delta\tau)] + \text{Im}[F^{T}(\tau-\Delta\tau)] \right\} \right)}{\exp{\left(  \frac{I_{+}}{e^*} \left\{ \text{Re}[F^{1}(\Delta\tau)] + \text{Re}[F^{2}(\Delta\tau)] + \text{Re}[F^{T}(\tau-\Delta\tau)] \right\} \right)}} 
\right.
\nonumber\\
& + \left. 
 \frac{
\Theta(\Delta\tau - \tau)
\cos \left( \frac{I_{-}}{e^*}\left\{ \text{Im}[F^{1} (\tau)] + \text{Im}[F^{2}(\tau)] \right\} \right)}
{\exp{\left(\frac{I_{+}}{e^*}\left\{ \text{Re}[F^{1}(\tau)] + \text{Re}[F^{2}(\tau)] \right\} \right)}}
\right].
\label{Eq:Final_T_noise2}
\end{align}       
\end{widetext}

For Lorentzian-shaped anyon pulses, we have
\begin{align}
F^{i}(\tau) =& \int_{-\infty}^{\infty} du \left( 1-e^{i2\tilde{g}_{A,i}\tilde{g}_{A,i}\left[ \text{tan}^{-1}\left(\frac{u}{\tau_{w}}\right) - \text{tan}^{-1}\left(\frac{u-\tau}{\tau_{w}} \right) \right]} \right) \nonumber\\
F^{T}(\tau) =& \int_{-\infty}^{\infty} du \left( 1-e^{i2 \bm{g_{A}}.\bm{g_{A}}\left[ \text{tan}^{-1}\left(\frac{u}{\tau_{w}}\right) - \text{tan}^{-1}\left(\frac{u-\tau}{\tau_{w}}\right) \right] }\right)
\end{align}
where $\tau_{w}$ is the width of the injected anyon pulse in the temporal domain (assumed to be the same for right and left moving edge modes). Note that in the limit $\tau_{w} \longrightarrow 0$, the injected anyon pulse becomes a delta-shaped pulse and these functions reduce to $F^{i}(\tau) = \tau (1-e^{-2i\pi\tilde{g}_{A,i}\tilde{g}_{A,i}})$ and $F^{T}(\tau) = \tau (1-e^{-2i\pi\bm{g}_{A}\bm{g}_{A}})$.

\end{document}


\title{Supplementary material for ``Effect of inter-edge interaction in a Quantum Hall Collider"}
\author{Amulya Ratnakar}
\affiliation{Aix Marseille Univ, Université de Toulon, CNRS, CPT, Marseille, France}
\author{Flavio Ronetti}
\affiliation{Aix Marseille Univ, Université de Toulon, CNRS, CPT, Marseille, France}
\author{Benoît Grémaud}
\affiliation{Aix Marseille Univ, Université de Toulon, CNRS, CPT, Marseille, France}
\author{Laurent Raymond}
\affiliation{Aix Marseille Univ, Université de Toulon, CNRS, CPT, Marseille, France}
\author{Thierry Martin}
\affiliation{Aix Marseille Univ, Université de Toulon, CNRS, CPT, Marseille, France}
\author{Thibaut Jonckheere}
\affiliation{Aix Marseille Univ, Université de Toulon, CNRS, CPT, Marseille, France}
\author{Jérôme Rech}
\affiliation{Aix Marseille Univ, Université de Toulon, CNRS, CPT, Marseille, France}

\maketitle
\section{2/5 edge QH collider setup}
Consider a Quantum Hall (QH) collider geometry with the bulk filling fraction $2/5$ and an edge structure as described in the main text. Also, consider the short-range inter-edge Coulomb interactions between the two co-propagating edge modes of $1/3$ and $1/15$ conductance. The collider consists of two source QPCs ($S_{1}$ and $S_{2}$) placed at $-L$ and $L$ from the collider QPC, respectively. The central QPC is placed at $x=0$. The QPC $S_{1}$ ($S_{2}$) is biased such that it injects the two most relevant anyonic excitations (as described in the main text) arbitrarily on the upper (lower) edge. The prepared state $\ket{\varphi_{\lbrace \gamma^{U}_{i},\gamma^L_{j} \rbrace}}$ is given by
\begin{eqnarray}
\ket{\varphi_{\lbrace \gamma^{U}_{i},\gamma^L_{j} \rbrace}} &=& \prod_{i=1}^{N_{U}}\prod_{j=1}^{N_L} \psi^{U\dagger}_{\gamma_{N^{U} -i+1}}(-L,-T^{U}_{N^{U}-i+1}) \psi^{L\dagger}_{\gamma_{N^L -j+1}}(L,-T^L_{N^L-j+1}) \ket{0},
\end{eqnarray}
where $N^{U/L}$ is the number of anyons injected in the upper/lower edge above the ground state $\ket{0}$, at time $-T^{U/L}_{N}<-T^{U/L}_{N-1}<....<-T^{U/L}_{1}$. The $\gamma^{U/L}_{i}$ denotes the type of anyon (among the two possible cases) injected at a time $T^{U/L}_{i}$ in the upper/lower edge. The corresponding density matrix, associated with the injection of $N^{U}+N^L$ anyons, is given by
\begin{eqnarray}
    \hat{\rho} = \sum_{\lbrace\gamma^{U}_{i},\gamma^L_{i}\rbrace}P_{\lbrace \gamma^{U}_{i},\gamma^L_{i}\rbrace} \ket{\varphi_{\lbrace \gamma^{U}_{i},\gamma^L_{j} \rbrace}}\bra{\varphi_{\lbrace \gamma^{U}_{i},\gamma^L_{j} \rbrace}}
\end{eqnarray}
where $P_{\lbrace \gamma^{U}_{i},\gamma^L_{i}\rbrace}$ is the probability of having the system in the $N^{U}+N^L$-anyon state of $\ket{\varphi_{\lbrace \gamma^{U}_{i},\gamma^L_{j} \rbrace}}$. The expectation value of any time-dependent observable $\hat{O}(t)$ is then given by
\begin{eqnarray}
    \langle \hat{O}(t) \rangle &=& Tr[\hat{O}(t)\hat{\rho}] \nonumber\\
    &=& \sum_{\lbrace\gamma^{U}_{i},\gamma^L_{i}\rbrace}P_{\lbrace \gamma^{U}_{i},\gamma^L_{i}\rbrace}\bra{\varphi_{\lbrace \gamma^{U}_{i},\gamma^L_{j} \rbrace}} \hat{O}(t)\ket{\varphi_{\lbrace \gamma^{U}_{i},\gamma^L_{j} \rbrace}}
\end{eqnarray}

The arbitrary quasi-hole and quasi-particle excitation takes place at the central QPC and is taken care of by the tunneling  term $H_{T}$ in the Hamiltonian, such that,
\begin{eqnarray}
    H_{T}(t) &=& \sum_{\alpha} \left(\Gamma_{\alpha} \psi^{U\dagger}_{\alpha}(0,t)\psi^{L}_{\alpha}(0,t) + H.C \right) \nonumber\\
    &=& \sum_{\alpha} \left(\Gamma_{\alpha} e^{-i\vec{g_{\alpha}}.\vec{\phi}^{U}(0,t)+i\vec{g_{\alpha}}.\vec{\phi}^{L}(0,t)} + H.C \right) \nonumber\\
    &=& \sum_{\alpha}\sum_{\xi = \pm} \left( \Gamma_{\alpha}e^{-i \vec{g_{\alpha}}.\vec{\phi}^{U}(0,t) + i\vec{g_{\alpha}}.\vec{\phi}^{L}(0,t)} \right)^{\xi}\\
    I_{T}(t) &=& i\sum_{\alpha}\sum_{\xi=\pm}\xi q_{\alpha}  \left( \Gamma_{\alpha} e^{-i \vec{g_{\alpha}}.\vec{\phi}^{U}(0,t) + i\vec{g_{\alpha}}.\vec{\phi}^{L}(0,t)} \right)^{\xi}.
\end{eqnarray}

Now the expectation value of the tunneling current $\langle \hat{I}_{T}(t)\rangle$ can be calculated using the Keldysh technique, yielding
\begin{eqnarray}
\langle \hat{I}_{T}(t)\rangle &=& \frac{1}{2}\sum_{\eta=\pm}\sum_{\lbrace\gamma^{U}_{i},\gamma^L_{i}\rbrace}P_{\lbrace \gamma^{U}_{i},\gamma^L_{i}\rbrace}\bra{\varphi_{\lbrace \gamma^{U}_{i},\gamma^L_{j} \rbrace}} \hat{T}_{K} \hat{I}_{T}(t^{\eta}) e^{-i\int_{k}H_{T}(t')dt'}\ket{\varphi_{\lbrace \gamma^{U}_{i},\gamma^L_{j} \rbrace}} \nonumber\\
&=& \frac{1}{2}\sum_{\eta=\pm}\sum_{\lbrace\gamma^{U}_{i},\gamma^L_{i}\rbrace}P_{\lbrace \gamma^{U}_{i},\gamma^L_{i}\rbrace}\bra{\varphi_{\lbrace \gamma^{U}_{i},\gamma^L_{j} \rbrace}} \hat{T}_{K} \hat{I}_{T}(t^{\eta})\left(1-i\sum_{\eta'=\pm}\eta'\int_{-\infty}^{\infty}H_{T}(t^{\eta'}_{1})dt_{1}\right)\ket{\varphi_{\lbrace \gamma^{U}_{i},\gamma^L_{j} \rbrace}} \nonumber\\
&=& -\frac{i}{2}\sum_{\eta\eta'=\pm}\eta'\sum_{\lbrace\gamma^{U}_{i},\gamma^L_{i}\rbrace}P_{\lbrace \gamma^{U}_{i},\gamma^L_{i}\rbrace}\int_{-\infty}^{\infty}dt_{1}\bra{\varphi_{\lbrace \gamma^{U}_{i},\gamma^L_{j} \rbrace}} \hat{T}_{K} \hat{I}_{T}(t^{\eta})H_{T}(t^{\eta'}_{1})\ket{\varphi_{\lbrace \gamma^{U}_{i},\gamma^L_{j} \rbrace}}
\end{eqnarray}

The tunneling noise is defined as 
\begin{eqnarray}
\langle S(t)\rangle = \int dt' \langle \delta \hat{I}(t)\delta \hat{I}(t') \rangle   
\end{eqnarray}
where $\delta I(t) = \hat{I}(t)-\langle \hat{I}(t) \rangle$. To the lowest order in the tunneling amplitude $\Gamma$, the noise is given by $S(t) = \int dt'\langle \hat{I}(t)\hat{I}(t') \rangle $.

As such, the tunneling noise at the central QPC is given by
\begin{eqnarray}
\langle S(t)\rangle &=& \frac{1}{2}\int dt' \sum_{\eta \eta'}\sum_{\lbrace \gamma^{U}_{i},\gamma^L_{j}\rbrace} P_{\lbrace \gamma^{U}_{i},\gamma^L_{j}\rbrace} \bra{\varphi_{\lbrace \gamma^U_{j},\gamma^L_{j}\rbrace}} \hat{I}_{T}(t^{\eta})\hat{I}_{T}(t'^{\eta'})e^{-i \int_{k} dt_1 H_{T}(t_1)} \ket{\varphi_{\lbrace \gamma^U_{j},\gamma^L_{j}\rbrace}} \nonumber\\
&=& \frac{1}{2}\int dt' \sum_{\eta \eta'}\sum_{\lbrace \gamma^{U}_{i},\gamma^L_{j}\rbrace} P_{\lbrace \gamma^{U}_{i},\gamma^L_{j}\rbrace} \bra{\varphi_{\lbrace \gamma^U_{j},\gamma^L_{j}\rbrace}} \hat{T}_{K}\lbrace\hat{I}_{T}(t^{\eta})\hat{I}_{T}(t'^{\eta'})\rbrace \ket{\varphi_{\lbrace \gamma^U_{j},\gamma^L_{j}\rbrace}} \nonumber\\
&=& -\frac{1}{2}\sum_{\alpha\alpha'}\sum_{\xi\xi'}\sum_{\eta\eta'} q_{\alpha}q_{\alpha'}\xi\xi' \Gamma^{\xi}_{\alpha} \Gamma^{\xi'}_{\alpha'}\sum_{\lbrace \gamma^{U}_{i},\gamma^L_{j}\rbrace}P_{\lbrace \gamma^{U}_{i},\gamma^L_{j}\rbrace} \int dt' \nonumber\\
&& \bra{\varphi_{\lbrace \gamma^U_{j},\gamma^L_{j}\rbrace}} \hat{T}_{K}\lbrace e^{-i\xi\vec{g}_{\alpha}.\vec{\phi}^{U}(0,t^{\eta})+i\xi\vec{g}_{\alpha}.\vec{\phi}^L(0,t^{\eta})} e^{-i\xi'\vec{g}_{\alpha'}.\vec{\phi}^{U}(0,t'^{\eta'})+i\xi'\vec{g}_{\alpha'}.\vec{\phi}^L(0,t'^{\eta'})} \rbrace
\ket{\varphi_{\lbrace \gamma^U_{j},\gamma^L_{j}\rbrace}}
\label{Eq:Noise_def}
\end{eqnarray}

We first calculate the tunneling noise using the Keldysh formalism. The above equation, Eq.~\eqref{Eq:Noise_def}, is non-zero only when $\alpha = \alpha'$ and $\xi' = -\xi$. As a result
\begin{eqnarray}
\langle S(t)\rangle &=& \frac{1}{2}\sum_{\alpha}\sum_{\xi}\sum_{\eta\eta'} q_{\alpha}^{2} |\Gamma_{\alpha}|^{2} \sum_{\lbrace \gamma^{U}_{i},\gamma^L_{j}\rbrace}P_{\lbrace \gamma^{U}_{i},\gamma^L_{j}\rbrace} \int dt' \nonumber\\
&& \bra{\varphi_{\lbrace \gamma^U_{j},\gamma^L_{j}\rbrace}} \hat{T}_{K}\lbrace e^{-i\xi\vec{g}_{\alpha}.\vec{\phi}^{U}(0,t^{\eta})+i\xi\vec{g}_{\alpha}.\vec{\phi}^L(0,t^{\eta})} e^{i\xi\vec{g}_{\alpha}.\vec{\phi}^{U}(0,t'^{\eta'})-i\xi\vec{g}_{\alpha}.\vec{\phi}^L(0,t'^{\eta'})} \rbrace
\ket{\varphi_{\lbrace \gamma^U_{j},\gamma^L_{j}\rbrace}}
\label{Eq:Noise_1}
\end{eqnarray}
Hence, the expectation value in Eq.~\eqref{Eq:Noise_1} is given by
\begin{eqnarray}
\bra{\varphi_{\lbrace \gamma^U_{i},\gamma^L_{i} \rbrace}}....\ket{\varphi_{\lbrace \gamma^U_{i},\gamma^L_{i} \rbrace}} &=& \bra{0}\prod_{i=1}^{N_{U}}\prod_{j=1}^{N_L} \psi^L_{\gamma^L_{j}}(L,-T^{\eta_{2}}_{j}) \psi^U_{\gamma^U_{i}}(-L,-T^{\eta_{2}}_{i})\nonumber\\
&& \hat{T}_{k}\lbrace e^{-i\xi\vec{g}_{\alpha}.\vec{\phi}^{U}(0,t^{\eta})+i\xi\vec{g}_{\alpha}.\vec{\phi}^L(0,t^{\eta})} e^{i\xi\vec{g}_{\alpha}.\vec{\phi}^{U}(0,t'^{\eta'})-i\xi\vec{g}_{\alpha}.\vec{\phi}^L(0,t'^{\eta'})} \rbrace\nonumber\\
&& \prod_{i=1}^{N_{U}}\prod_{j=1}^{N_L} \psi^{U\dagger}_{\gamma_{N_{U} -i+1}}(-L,-T_{N_{U}-i+1}) \psi^{L\dagger}_{\gamma_{N_L -j+1}}(L,-T_{N_L-j+1}) \ket{0}
\end{eqnarray}

\begin{eqnarray}
\bra{\varphi_{\lbrace \gamma^U_{i},\gamma^L_{i} \rbrace}}....\ket{\varphi_{\lbrace \gamma^U_{i},\gamma^L_{i} \rbrace}} &=& \bra{0}\prod_{i=1}^{N_{U}}\prod_{j=1}^{N_L}e^{i \vec{g}_{\gamma^L_{j}}.\vec{\phi}^L(L,-T^{\eta_{2}}_{j}) + i\vec{g}_{\gamma^U_{i}}.\vec{\phi}^U(-L,-T^{\eta_{2}}_{i})}\nonumber\\
&& \hat{T}_{K}\lbrace e^{-i\xi\vec{g}_{\alpha}.\vec{\phi}^{U}(0,t^{\eta})+i\xi\vec{g}_{\alpha}.\vec{\phi}^L(0,t^{\eta})} e^{i\xi\vec{g}_{\alpha}.\vec{\phi}^{U}(0,t'^{\eta'})-i\xi\vec{g}_{\alpha}.\vec{\phi}^L(0,t'^{\eta'})} \rbrace \nonumber\\
&& \prod_{i=1}^{N_{U}}\prod_{j=1}^{N_L} e^{-i \vec{g}_{\gamma^U_{N_{U}-i+1}}.\vec{\phi}^U(-L,-T^{\eta_{1}}_{N_{U}-i+1}) - i\vec{g}_{\gamma^L_{N_L-j+1}}.\vec{\phi}^L(L,-T^{\eta_{1}}_{N_L-j+1})} \ket{0} \nonumber\\
&=& \bra{0}\prod_{i=1}^{N_{U}}\prod_{j=1}^{N_L}e^{i \tilde{g}_{\gamma^L_{j}}.\tilde{\phi}^L(L,-T^{\eta_{2}}_{j}) + i\tilde{g}_{\gamma^U_{i}}.\tilde{\phi}^U(-L,-T^{\eta_{2}}_{i})}\nonumber\\
&& \hat{T}_{K}\lbrace e^{-i\xi\tilde{g}_{\alpha}.\tilde{\phi}^{U}(0,t^{\eta})+i\xi\tilde{g}_{\alpha}.\tilde{\phi}^L(0,t^{\eta})} e^{i\xi\tilde{g}_{\alpha}.\tilde{\phi}^{U}(0,t'^{\eta'})-i\xi\tilde{g}_{\alpha}.\tilde{\phi}^L(0,t'^{\eta'})} \rbrace \nonumber\\
&& \prod_{i=1}^{N_{U}}\prod_{j=1}^{N_L} e^{-i \tilde{g}_{\gamma^U_{N_{U}-i+1}}.\tilde{\phi}^U(-L,-T^{\eta_{1}}_{N_{U}-i+1}) - i\tilde{g}_{\gamma^L_{N_L-j+1}}.\tilde{\phi}^L(L,-T^{\eta_{1}}_{N_L-j+1})} \ket{0}
\end{eqnarray}
Now the effect of inter-edge interaction can be taken into account by expressing the interacting bosonic modes $\vec{\phi}^{U/L}(x,t)$ as the linear combination of free Bogoliubov modes $\tilde{\phi}(x,t)$, as $\vec{\phi}^{U/L}(x,t) = M \tilde{\phi}(x,t)$, where $M$ is the diagonalizing matrix. Then the correlation above can be written as
\begin{eqnarray}  
\bra{\varphi_{\lbrace \gamma^U_{i},\gamma^L_{i} \rbrace}}....\ket{\varphi_{\lbrace \gamma^U_{i},\gamma^L_{i} \rbrace}}&=& \bra{0}\prod_{i=1}^{N_{U}}\prod_{j=1}^{N_L}e^{i \sum_{k=1}^{2}\tilde{g}^{k}_{\gamma^L_{j}}\tilde{\phi}^L_{k}(L,-T^{\eta_{2}}_{j}) + i\sum_{k=1}^{2}\tilde{g}^{k}_{\gamma^U_{i}}\tilde{\phi}^U_{k}(-L,-T^{\eta_{2}}_{i})}\nonumber\\
&& \hat{T}_{K}\lbrace e^{-i\xi\sum_{k=1}^{2}\tilde{g}^{k}_{\alpha}\tilde{\phi}^{U}_{k}(0,t^{\eta})+i\xi\sum_{k=1}^{2}\tilde{g}^{k}_{\alpha}\tilde{\phi}^L_{k}(0,t^{\eta})} e^{i\xi\sum_{k=1}^{2}\tilde{g}^{k}_{\alpha}\tilde{\phi}^{U}_{k}(0,t'^{\eta'})-i\xi\sum_{k=1}^{2}\tilde{g}^{k}_{\alpha}\tilde{\phi}^L_{k}(0,t'^{\eta'})} \rbrace \nonumber\\
&& \prod_{i=1}^{N_{U}}\prod_{j=1}^{N_L} e^{-i\sum_{k=1}^{2} \tilde{g}^{k}_{\gamma^U_{N_{U}-i+1}}\tilde{\phi}^U_{k}(-L,-T^{\eta_{1}}_{N_{U}-i+1}) - i\sum_{k=1}^{2}\tilde{g}^{k}_{\gamma^L_{N_L-j+1}}\tilde{\phi}^L_{k}(L,-T^{\eta_{1}}_{N_L-j+1})} \ket{0}\nonumber\\
&=& \prod_{k=1}^{2}\left[\bra{0}\prod_{i=1}^{N_{U}}\prod_{j=1}^{N_L}e^{i \tilde{g}^{k}_{\gamma^L_{j}}\tilde{\phi}^L_{k}(L,-T^{\eta_{2}}_{j}) + i\tilde{g}^{k}_{\gamma^U_{i}}\tilde{\phi}^U_{k}(-L,-T^{\eta_{2}}_{i})}\right.\nonumber\\
&& \left.\hat{T}_{K}\lbrace e^{-i\xi\tilde{g}^{k}_{\alpha}\tilde{\phi}^{U}_{k}(0,t^{\eta})+i\xi\tilde{g}^{k}_{\alpha}\tilde{\phi}^L_{k}(0,t^{\eta})} e^{i\xi\tilde{g}^{k}_{\alpha}\tilde{\phi}^{U}_{k}(0,t'^{\eta'})-i\xi\tilde{g}^{k}_{\alpha}\tilde{\phi}^L_{k}(0,t'^{\eta'})} \rbrace \right.\nonumber\\
&& \left.\prod_{i=1}^{N_{U}}\prod_{j=1}^{N_L} e^{-i \tilde{g}^{k}_{\gamma^U_{N_{U}-i+1}}\tilde{\phi}^U_{k}(-L,-T^{\eta_{1}}_{N_{U}-i+1}) - i\tilde{g}^{k}_{\gamma^L_{N_L-j+1}}\tilde{\phi}^L_{k}(L,-T^{\eta_{1}}_{N_L-j+1})} \ket{0} \right] \nonumber\\
&=& \prod_{k=1}^{2}\left[\bra{0}\prod_{i=1}^{N_{U}}e^{i\tilde{g}^{k}_{\gamma^U_{i}}\tilde{\phi}^U_{k}(-L,-T^{\eta_{2}}_{i})}\hat{T}_{K}\lbrace e^{-i\xi\tilde{g}^{k}_{\alpha}\tilde{\phi}^{U}_{k}(0,t^{\eta})}e^{i\xi\tilde{g}^{k}_{\alpha}\tilde{\phi}^{U}_{k}(0,t'^{\eta'})}\rbrace\right.\nonumber\\
&&\left.\prod_{i=1}^{N_{U}}e^{-i \tilde{g}^{k}_{\gamma^U_{N_{U}-i+1}}\tilde{\phi}^U_{k}(-L,-T^{\eta_{1}}_{N_{U}-i+1})}\ket{0} \times \bra{0}\prod_{j=1}^{N_L}e^{i \tilde{g}^{k}_{\gamma^L_{j}}\tilde{\phi}^L_{k}(L,-T^{\eta_{2}}_{j})}\right.\nonumber\\
&&\left.  \hat{T}_{K}\lbrace e^{i\xi\tilde{g}^{k}_{\alpha}\tilde{\phi}^L_{k}(0,t^{\eta})} e^{-i\xi\tilde{g}^{k}_{\alpha}\tilde{\phi}^L_{k}(0,t'^{\eta'})} \rbrace\prod_{j=1}^{N_L} e^{ - i\tilde{g}^{k}_{\gamma^L_{N_L-j+1}}\tilde{\phi}^L_{k}(L,-T^{\eta_{1}}_{N_L-j+1})} \ket{0} \right]
\end{eqnarray}

From this we can express, 
\begin{eqnarray}
S^{U}_{k} &=&   \bra{0}\prod_{i=1}^{N_{U}}e^{i\tilde{g}^{k}_{\gamma^U_{i}}\tilde{\phi}^U_{k}(-L,-T^{\eta_{2}}_{i})}\hat{T}_{K}\lbrace e^{-i\xi\tilde{g}^{k}_{\alpha}\tilde{\phi}^{U}_{k}(0,t^{\eta})}e^{i\xi\tilde{g}^{k}_{\alpha}\tilde{\phi}^{U}_{k}(0,t'^{\eta'})}\rbrace \prod_{i=1}^{N_{U}}e^{-i \tilde{g}^{k}_{\gamma^U_{N_{U}-i+1}}\tilde{\phi}^U_{k}(-L,-T^{\eta_{1}}_{N_{U}-i+1})}\ket{0}\nonumber\\
&=& \bra{0}\hat{T}_{K}\lbrace\prod_{i=1}^{N_{U}}e^{i\tilde{g}^{k}_{\gamma^U_{i}}\tilde{\phi}^U_{k}(-L,-T^{\eta_{2}}_{i})} e^{-i\xi\tilde{g}^{k}_{\alpha}\tilde{\phi}^{U}_{k}(0,t^{\eta})}e^{i\xi\tilde{g}^{k}_{\alpha}\tilde{\phi}^{U}_{k}(0,t'^{\eta'})} \prod_{i=1}^{N_{U}}e^{-i \tilde{g}^{k}_{\gamma^U_{N_{U}-i+1}}\tilde{\phi}^U_{k}(-L,-T^{\eta_{1}}_{N_{U}-i+1})}\rbrace\ket{0}
\end{eqnarray}
and,
\begin{eqnarray}
S^L_{k} &=& \bra{0}\prod_{j=1}^{N_L}e^{i \tilde{g}^{k}_{\gamma^L_{j}}\tilde{\phi}^L_{k}(L,-T^{\eta_{2}}_{j})}\hat{T}_{K}\lbrace e^{i\xi\tilde{g}^{k}_{\alpha}\tilde{\phi}^L_{k}(0,t^{\eta})} e^{-i\xi\tilde{g}^{k}_{\alpha}\tilde{\phi}^L_{k}(0,t'^{\eta'})} \rbrace\prod_{j=1}^{N_L} e^{ - i\tilde{g}^{k}_{\gamma^L_{N_L-j+1}}\tilde{\phi}^L_{k}(L,-T^{\eta_{1}}_{N_L-j+1})} \ket{0}\nonumber\\
&=& \bra{0}\hat{T}_{K}\lbrace\prod_{j=1}^{N_L}e^{i \tilde{g}^{k}_{\gamma^L_{j}}\tilde{\phi}^L_{k}(L,-T^{\eta_{2}}_{j})} e^{i\xi\tilde{g}^{k}_{\alpha}\tilde{\phi}^L_{k}(0,t^{\eta})} e^{-i\xi\tilde{g}^{k}_{\alpha}\tilde{\phi}^L_{k}(0,t'^{\eta'})} \prod_{j=1}^{N_L} e^{ - i\tilde{g}^{k}_{\gamma^L_{N_L-j+1}}\tilde{\phi}^L_{k}(L,-T^{\eta_{1}}_{N_L-j+1})} \rbrace\ket{0}.
\end{eqnarray}
Here, we consider four Keldysh time contours, denoted by the Keldysh indices $\eta_{1},\eta_{2},\eta,\eta'$. Any time on the contour $\eta_{2}$ is greater than any time on other contours, $\eta,\eta',\eta_{1}$. Similarly, any time on the contour $\eta_{1}$ is less than any time on other contours$\eta,\eta',\eta_{2}$. Now, using the formula
\begin{eqnarray}
\bra{0}\hat{T}_{k}\lbrace\prod_{j}e^{i\alpha_{j}\phi(x_{j},t^{\eta_{j}}_{j})}\rbrace\ket{0} &=& \mathrm{Exp}[-\sum_{i<j}\alpha_{i}\alpha_{j}G^{\eta_{i}\eta_{j}}(x_{i}-x_{j},t_{i}-t_{j})],   
\end{eqnarray}
we can calculate $S^{U/L}_{k}$, as 
\begin{eqnarray}
S^{U}_{k} &=& (\tilde{G}^{\eta\eta'}_{k}(t-t'))^{(\tilde{g}^{k}_{\alpha})^{2}}e^{2i\pi\xi\sum_{i=1}^{N_{u}}\tilde{g}^{k}_{\alpha}\tilde{g}^{k}_{\gamma^U_{i}}(\Theta(T_{i}-\frac{L}{\tilde{v}_{k}}+t)-\Theta(T_{i}-\frac{L}{\tilde{v}_{k}}+t'))} \nonumber\\
S^L_{k} &=& (\tilde{G}^{\eta\eta'}_{k}(t-t'))^{(\tilde{g}^{k}_{\alpha})^{2}}e^{-2i\pi\xi\sum_{j=1}^{N_L}\tilde{g}^{k}_{\alpha}\tilde{g}^{k}_{\gamma^L_{j}}(\Theta(T_{j}-\frac{L}{\tilde{v}_{k}}+t)-\Theta(T_{j}-\frac{L}{\tilde{v}_{k}}+t'))} 
\end{eqnarray}
up to a normalization factor, given by $\langle{\varphi_{\lbrace \gamma^U_{i},\gamma^L_{i} \rbrace}}|\varphi_{\lbrace \gamma^U_{i},\gamma^L_{i} \rbrace}\rangle$. As a result, the noise can be written as
\begin{eqnarray}
\langle S(t)\rangle &=& \frac{1}{2}\sum_{\alpha}\sum_{\xi}\sum_{\eta\eta'} q_{\alpha}^{2} |\Gamma_{\alpha}|^{2} \sum_{\lbrace \gamma^{U}_{i},\gamma^L_{j}\rbrace}P_{\lbrace \gamma^{U}_{i},\gamma^L_{j}\rbrace} \int_{-\infty}^{\infty} dt'\left[ \prod_{k=1}^{2}(\tilde{G}^{\eta\eta'}_{k}(t-t')
)^{2(\tilde{g}^{k}_{\alpha})^{2}} \right] \nonumber\\
&& \times e^{2\pi i \xi(\Lambda^{U}(t,t',\lbrace T^{U}_{i}\rbrace)-\Lambda^L(t,t',\lbrace T^L_{j}\rbrace))},
\end{eqnarray}
where the time-dependent phases are defined as $\Lambda^{U}(t,t',\lbrace T^{U}_{i}\rbrace) = \sum_{i=1}^{N_{U}}\sum_{k=1}^{2}\tilde{g}^{k}_{\alpha}\tilde{g}^{k}_{\gamma^U_{i}}(\Theta(T^{U}_{i}-\frac{L}{\tilde{v}_{k}}+t)-\Theta(T^{U}_{i}-\frac{L}{\tilde{v}_{k}}+t'))$ and $\Lambda^L(t,t',\lbrace T^L_{j}\rbrace) = \sum_{j=1}^{N_L}\sum_{k=1}^{2}\tilde{g}^{k}_{\alpha}\tilde{g}^{k}_{\gamma^L_{j}}(\Theta(T^L_{j}-\frac{L}{\tilde{v}_{k}}+t)-\Theta(T^L_{j}-\frac{L}{\tilde{v}_{k}}+t'))$. $T^{U/L}_{i}$ is the time of injecting $\gamma^{u/d}_{i}$-type  quasi-particle from source QPC $S_{1/2}$. The Green's function $\tilde{G}^{\eta\eta'}_{k}(t-t,)$ is given by
\begin{eqnarray}
\tilde{G}^{\eta\eta'}_{k}(x,t) &=& \frac{\sinh(\frac{i\pi\alpha}{\tilde{v}_{k}\beta})}{\sinh(\frac{\pi}{\tilde{v}_{k}\beta}(\sigma_{\eta\eta'}(t)(x-\tilde{v}_{k}t))+i\alpha)};\nonumber\\
\sigma_{\eta\eta'}(t)&=& \frac{1}{2}((\eta - \eta')+\mathrm{Sgn}(t)(\eta+\eta')).
\end{eqnarray}
Using the fact that $\sum_{\eta\eta'}\prod_{k=1}^{2}(\tilde{G}^{\eta\eta'}_{k}(t-t'))^{2(\tilde{g}^{k}_{\alpha})^{2}} = 2\Theta(t-t')(\prod_{k=1}^{2}(\tilde{G}^{-+}_{k}(t-t'))^{2(\tilde{g}^{k}_{\alpha})^{2}} + \prod_{k=1}^{2}(\tilde{G}^{-+}_{k}(t'-t))^{2(\tilde{g}^{k}_{\alpha})^{2}})$, we can express the tunneling noise as
\begin{eqnarray}
\langle S(t)\rangle &=& \sum_{\alpha}\sum_{\xi} q_{\alpha}^{2} |\Gamma_{\alpha}|^{2} \sum_{\lbrace \gamma^{U}_{i},\gamma^L_{j}\rbrace}P_{\lbrace \gamma^{U}_{i},\gamma^L_{j}\rbrace} \int_{-\infty}^{t} dt'\left[ \prod_{k=1}^{2}(\tilde{G}^{-+}_{k}(t-t'))^{2(\tilde{g}^{k}_{\alpha})^{2}} + \prod_{k=1}^{2}(\tilde{G}^{-+}_{k}(t'-t))^{2(\tilde{g}^{k}_{\alpha})^{2}} \right] \nonumber\\
&& \times e^{2\pi i \xi(\Lambda^{U}(t,t',\lbrace T^{U}_{i}\rbrace)-\Lambda^L(t,t',\lbrace T^L_{j}\rbrace))}\nonumber\\
&=& 2\sum_{\alpha}q_{\alpha}^{2} |\Gamma_{\alpha}|^{2}\sum_{\lbrace \gamma^{U}_{i},\gamma^L_{j}\rbrace}P_{\lbrace \gamma^{U}_{i},\gamma^L_{j}\rbrace} \int_{-\infty}^{t} dt'\left[ \prod_{k=1}^{2}(\tilde{G}^{-+}_{k}(t-t'))^{2(\tilde{g}^{k}_{\alpha})^{2}} + \prod_{k=1}^{2}(\tilde{G}^{-+}_{k}(t'-t))^{2(\tilde{g}^{k}_{\alpha})^{2}} \right] \nonumber\\
&& \times \cos(2\pi (\Lambda^{U}(t,t',\lbrace T^{U}_{i}\rbrace)-\Lambda^L(t,t',\lbrace T^L_{j}\rbrace))).
\end{eqnarray}

Similarly to the noise calculation for two source QPCs, one can attempt a calculation for the expectation value of the tunneling current with two source QPCs. The latter is given by
\begin{eqnarray}
\langle I_{T}(t) \rangle &=& \sum_{\alpha}q_{\alpha} |\Gamma_{\alpha}|^{2}\sum_{\xi=\pm}\xi\sum_{\lbrace \gamma^U_{i},\gamma^L_{j}\rbrace}P_{\lbrace\gamma^U_{i},\gamma^L_{j}\rbrace}\nonumber\\
&& \int_{-\infty}^{t}dt_{1}  \left[\prod_{j=1}^{2} \left(G^{-+}_{j}(t-t_1)\right)^{2(\tilde{g}^{j}_{\alpha})^{2}} - \prod_{j=1}^{2} \left(G^{-+}_{j}(t_1-t)\right)^{2(\tilde{g}^{j}_{\alpha})^{2}}\right]e^{2i\pi\xi(\Lambda^{U}(t,t_{1},\lbrace T^{U}_{l} \rbrace)-\Lambda^L(t,t_{1},\lbrace T^L_{l} \rbrace))} \nonumber\\
&=& \sum_{\alpha}q_{\alpha} |\Gamma_{\alpha}|^{2}\sum_{\xi=\pm}\xi\sum_{\lbrace \gamma^U_{i},\gamma^L_{j}\rbrace}P_{\lbrace\gamma^U_{i},\gamma^L_{j}\rbrace}\int_{-\infty}^{t}dt_{1}  \left[\prod_{j=1}^{2} \left(G^{-+}_{j}(t-t_1)\right)^{2(\tilde{g}^{j}_{\alpha})^{2}} - \prod_{j=1}^{2} \left(G^{-+}_{j}(t_1-t)\right)^{2(\tilde{g}^{j}_{\alpha})^{2}}\right]\nonumber\\
&& \times \sin(2\pi (\Lambda^{U}(t,t',\lbrace T^{U}_{i}\rbrace)-\Lambda^L(t,t',\lbrace T^L_{j}\rbrace))).
\end{eqnarray}

\section{Poissonian Average}
Now, we perform a Poissonian average over the injected anyons. If the inter-edge interaction is present, then the edge velocities get renormalized. As such, the injected anyons split into excitations along the eigenmodes (with different renormalized velocities), reaching the central QPC at different times. For 2/5 QH edge structure, let the normalized velocities be given by $\tilde{v}_{1/2}$, such that $\tilde{v}_{1}>\tilde{v}_{2}$. Let $N^{U/L}_1$ be the number of anyonic excitations along the eigenmode with velocity $\tilde{v}_1$, reaching the central QPC between time $[t_1,t]$, such that $t_1<-T^{U/L}_l+\frac{L}{\tilde{v}_1}<t$ along the upper/lower edge. These excitations occurred at the source QPCs between time $t_1 - \frac{L}{\tilde{v}_1}<-T^{U/L}_l<t+\frac{L}{\tilde{v}_1}$. The probability of injecting $N^{U/L}_1$ anyons at the source QPCs between the time difference, $t-\frac{L}{\tilde{v}_1}-(t_1-\frac{L}{\tilde{v}_1}) = t-t_1$, is given by the Poissonian distribution given by $P_{N^{U/L}_1}(t-t_1)$. Since the velocity of the second eigenmode is $\tilde{v}_{2}<\tilde{v}_1$, excitations along the second eigenmode do not reach the central QPC in the same time interval and lag by the time interval $\Delta\tau = \frac{L}{\tilde{v}_2} - \frac{L}{\tilde{v}_1}$. To get the contribution of all the anyons reaching the central QPC in time $[t_1,t]$, we need to consider another set of $N^{U/L}_2$ anyons reaching the central QPC between time $[t_1,t_1+\Delta\tau]$, such that, $t_1<-T^{U/L}_l+\frac{L}{\tilde{v}_2}<t_1 + \Delta\tau$ and the probability of this Poissonian process is given by $P_{N_2}(\Delta\tau)$. Since two types of anyons can be injected at the source QPCs, let out of $N^{U/L}_1$, there are $m^{U/L}_1$ $A$-type anyons injected, and out of $N^{U/L}_2$, there are $m^{U/L}_2$ $B$-type anyons injected at the source QPC along the upper/lower edge. Since the injection processes are independent of each other, $P_{N^{U/L}_1}(t-t_1) = P_{m^{U/L}_1}(t-t_1)P_{N^{U/L}_1-m^{U/L}_1}(t-t_1)$ and $P_{N_2}(\Delta\tau) = P_{m^{U/L}_2}(\Delta\tau)P_{N^{U/L}_2-m^{U/L}_2}(\Delta\tau)$. The total probability of such a Poissonian injection process for which $t-t_1>\Delta\tau$ is given by
\begin{eqnarray}
P(\lbrace N^{U/L}_{1},m^{U/L}_{1},t-t'\rbrace; \lbrace N^{U/L}_{2},m^{U/L}_{2},\Delta\tau\rbrace) &=& P_{m^{U}_{1}}(t-t')P_{N^{U}_{1}-m^{U}_{1}}(t-t')P_{m^{U}_{2}}(\Delta\tau) P_{N^{U}_{2}-m^{U}_{2}}(\Delta\tau)\nonumber\\
&&\times P_{m^L_{1}}(t-t')P_{N^L_{1}-m^L_{1}}(t-t')P_{m^L_{2}}(\Delta\tau) P_{N^L_{2}-m^L_{2}}(\Delta\tau),
\end{eqnarray}
where
\begin{eqnarray}
P_{N}(t-t') &=& \frac{1}{N!}(\gamma(t-t'))^{N}e^{-\gamma(t-t')}.\nonumber    
\end{eqnarray}
The Poissonian averaging includes identifying the probability 
\begin{equation}
\sum_{\lbrace \gamma^{U}_{i},\gamma^L_{j} \rbrace} P_{\lbrace \gamma^{U}_{i},\gamma^L_{j} \rbrace} \longrightarrow \sum_{N^{U/L}_{1/2}=0}^{\infty}\sum_{m^{U/L}_{1/2}=0}^{N^{U/L}_{1/2}}P(\lbrace N^{U/L}_{1},m^{U/L}_{1},t-t'\rbrace; \lbrace N^{U/L}_{2},m^{U/L}_{2},\Delta\tau\rbrace).
\end{equation}
However, if $t-t'<\Delta\tau$, then the Poissonian injection process is determined by the probability
\begin{eqnarray}
P(\lbrace N^{U/L}_{1},m^{U/L}_{1},t-t'\rbrace; \lbrace N^{U/L}_{2},m^{U/L}_{2},t-t'\rbrace) &=& P_{m^{U}_{1}}(t-t')P_{N^{U}_{1}-m^{U}_{1}}(t-t')P_{m^{U}_{2}}(t-t') P_{N^{U}_{2}-m^{U}_{2}}(t-t')\times\nonumber\\
&& P_{m^L_{1}}(t-t')P_{N^L_{1}-m^L_{1}}(t-t')P_{m^L_{2}}(t-t') P_{N^L_{2}-m^L_{2}}(t-t').
\end{eqnarray}
\normalsize
The Poissonian averaged tunneling noise involves integrating over the injection time instant, weighted by the Poissonian probability for the event. The average over the anyon injection time is given by
\begin{eqnarray}
I^{U}_{1} &=& \int_{-t+\frac{L}{\tilde{v}_{1}}}^{-t'+\frac{L}{\tilde{v}_{1}}} \frac{D\lbrace T^{U}_{l} \rbrace}{(t-t')^{N^{U}_{1}}}e^{2\pi i \xi \sum_{k=1}^{2}\sum_{l=1}^{N^{U}_{1}}\tilde{g}^{k}_{\gamma^U_{l}}\tilde{g}^{k}_{\alpha}(\Theta(T^{U}_{l}-\frac{L}{\tilde{v}_{k}}+t)-\Theta(T^{U}_{l}-\frac{L}{\tilde{v}_{k}}+t'))}\nonumber\\
&=& \left[\int_{-t+\frac{L}{\tilde{v}_{1}}}^{-t'+\frac{L}{\tilde{v}_{1}}} \frac{d T^{U}_{l}}{t-t'} e^{2\pi i \xi \sum_{k=1}^{2} \tilde{g}^{k}_{1}\tilde{g}^{k}_{\alpha}(\Theta(T^{U}_{l}-\frac{L}{\tilde{v}_{k}}+t)-\Theta(T^{U}_{l}-\frac{L}{\tilde{v}_{k}}+t'))}\right]^{m^{U}_{1}}\nonumber\\
&& \times \left[\int_{-t+\frac{L}{\tilde{v}_{1}}}^{-t'+\frac{L}{\tilde{v}_{1}}} \frac{d T^{U}_{l}}{t-t'} e^{2\pi i \xi \sum_{k=1}^{2} \tilde{g}^{k}_{2}\tilde{g}^{k}_{\alpha}(\Theta(T^{U}_{l}-\frac{L}{\tilde{v}_{k}}+t)-\Theta(T^{U}_{l}-\frac{L}{\tilde{v}_{k}}+t'))}\right]^{N^{U}_{1}-m^{U}_{1}}\nonumber\\
I^{U}_{2} &=& \int_{-t'+\frac{L}{\tilde{v}_{1}}}^{-t'+\frac{L}{\tilde{v}_{2}}} \frac{D\lbrace T^{U}_{l} \rbrace}{(\Delta\tau)^{N^{U}_{2}}}e^{2\pi i \xi \sum_{k=1}^{2}\sum_{l=1}^{N^{U}_{2}}\tilde{g}^{k}_{\gamma^U_{l}}\tilde{g}^{k}_{\alpha}(\Theta(T^{U}_{l}-\frac{L}{\tilde{v}_{k}}+t)-\Theta(T^{U}_{l}-\frac{L}{\tilde{v}_{k}}+t'))}\nonumber\\
&=& \left[\int_{-t'+\frac{L}{\tilde{v}_{1}}}^{-t'+\frac{L}{\tilde{v}_{2}}} \frac{d T^{U}_{l}}{\Delta\tau} e^{2\pi i \xi \sum_{k=1}^{2} \tilde{g}^{k}_{1}\tilde{g}^{k}_{\alpha}(\Theta(T^{U}_{l}-\frac{L}{\tilde{v}_{k}}+t)-\Theta(T^{U}_{l}-\frac{L}{\tilde{v}_{k}}+t'))}\right]^{m^{U}_{2}}\nonumber\\
&& \times \left[\int_{-t'+\frac{L}{\tilde{v}_{1}}}^{-t'+\frac{L}{\tilde{v}_{2}}} \frac{d T^{U}_{l}}{\Delta\tau} e^{2\pi i \xi \sum_{k=1}^{2} \tilde{g}^{k}_{2}\tilde{g}^{k}_{\alpha}(\Theta(T^{U}_{l}-\frac{L}{\tilde{v}_{k}}+t)-\Theta(T^{U}_{l}-\frac{L}{\tilde{v}_{k}}+t'))}\right]^{N^{U}_{2}-m^{U}_{2}}.
\end{eqnarray}
Similarly,
\begin{eqnarray}
I^L_{1} &=& \int_{-t+\frac{L}{\tilde{v}_{1}}}^{-t'+\frac{L}{\tilde{v}_{1}}} \frac{D\lbrace T^L_{l} \rbrace}{(t-t')^{N^L_{1}}}e^{-2\pi i \xi \sum_{k=1}^{2}\sum_{l=1}^{N^L_{1}}\tilde{g}^{k}_{\gamma^L_{l}}\tilde{g}^{k}_{\alpha}(\Theta(T^L_{l}-\frac{L}{\tilde{v}_{k}}+t)-\Theta(T^L_{l}-\frac{L}{\tilde{v}_{k}}+t'))}\nonumber\\
&=& \left[\int_{-t+\frac{L}{\tilde{v}_{1}}}^{-t'+\frac{L}{\tilde{v}_{1}}} \frac{d T^L_{l}}{t-t'} e^{-2\pi i \xi \sum_{k=1}^{2} \tilde{g}^{k}_{1}\tilde{g}^{k}_{\alpha}(\Theta(T^L_{l}-\frac{L}{\tilde{v}_{k}}+t)-\Theta(T^L_{l}-\frac{L}{\tilde{v}_{k}}+t'))}\right]^{m^L_{1}}\nonumber\\
&& \times \left[\int_{-t+\frac{L}{\tilde{v}_{1}}}^{-t'+\frac{L}{\tilde{v}_{1}}} \frac{d T^L_{l}}{t-t'} e^{-2\pi i \xi \sum_{k=1}^{2} \tilde{g}^{k}_{2}\tilde{g}^{k}_{\alpha}(\Theta(T^L_{l}-\frac{L}{\tilde{v}_{k}}+t)-\Theta(T^L_{l}-\frac{L}{\tilde{v}_{k}}+t'))}\right]^{N^L_{1}-m^L_{1}}\nonumber\\
I^L_{2} &=& \int_{-t'+\frac{L}{\tilde{v}_{1}}}^{-t'+\frac{L}{\tilde{v}_{2}}} \frac{D\lbrace T^L_{l} \rbrace}{(\Delta\tau)^{N^L_{2}}}e^{-2\pi i \xi \sum_{k=1}^{2}\sum_{l=1}^{N^L_{2}}\tilde{g}^{k}_{\gamma^L_{l}}\tilde{g}^{k}_{\alpha}(\Theta(T^L_{l}-\frac{L}{\tilde{v}_{k}}+t)-\Theta(T^L_{l}-\frac{L}{\tilde{v}_{k}}+t'))}\nonumber\\
&=& \left[\int_{-t'+\frac{L}{\tilde{v}_{1}}}^{-t'+\frac{L}{\tilde{v}_{2}}} \frac{d T^L_{l}}{\Delta\tau} e^{-2\pi i \xi \sum_{k=1}^{2} \tilde{g}^{k}_{1}\tilde{g}^{k}_{\alpha}(\Theta(T^L_{l}-\frac{L}{\tilde{v}_{k}}+t)-\Theta(T^L_{l}-\frac{L}{\tilde{v}_{k}}+t'))}\right]^{m^L_{2}}\nonumber\\
&& \times \left[\int_{-t'+\frac{L}{\tilde{v}_{1}}}^{-t'+\frac{L}{\tilde{v}_{2}}} \frac{d T^L_{l}}{\Delta\tau} e^{-2\pi i \xi \sum_{k=1}^{2} \tilde{g}^{k}_{2}\tilde{g}^{k}_{\alpha}(\Theta(T^L_{l}-\frac{L}{\tilde{v}_{k}}+t)-\Theta(T^L_{l}-\frac{L}{\tilde{v}_{k}}+t'))}\right]^{N^L_{2}-m^L_{2}}   
\end{eqnarray}
with,
\begin{eqnarray}
f^{1 U/L}_{\alpha i}(\xi) &=& \int_{-t+\frac{L}{\tilde{v}_{1}}}^{-t'+\frac{L}{\tilde{v}_{1}}} \frac{d T^{U/L}_{l}}{t-t'} e^{\pm 2\pi i \xi \sum_{k=1}^{2} \tilde{g}^{k}_{1}\tilde{g}^{k}_{\alpha}(\Theta(T^L_{l}-\frac{L}{\tilde{v}_{k}}+t)-\Theta(T^L_{l}-\frac{L}{\tilde{v}_{k}}+t') )} \nonumber\\
&=& \begin{cases} 
      e^{\pm 2\pi i \xi \lambda^{1}_{\alpha i}} & t-t_1 < \Delta\tau \\
      \frac{e^{\pm 2\pi i \xi \lambda^{1}_{\alpha i}}}{t-t_1}\Delta\tau + e^{\pm 2\pi i \xi (\lambda^{1}_{\alpha i}+\lambda^{2}_{\alpha i})}\left(1-\frac{\Delta\tau}{t-t_1}\right) & t-t_1\geq \Delta\tau
   \end{cases} \nonumber\\
f^{2\xi}_{\alpha i}(\xi) &=& e^{2\pi i \xi \lambda^{2}_{\alpha i}}
\end{eqnarray}
However, in the limit of $\Delta\tau>t-t'$, the Poissonian averaging includes the weighted integrals  $I^{U/L}_{2}$ (with $P(\lbrace m_1,N_1-m_1,t-t'\rbrace;\lbrace m_2,N_2-m_2,t-t'\rbrace)$), which modifies to
\begin{eqnarray}
I^{U}_{2} &=& \int_{-t+\frac{L}{\tilde{v}_{2}}}^{-t'+\frac{L}{\tilde{v}_{2}}} \frac{D\lbrace T^{U}_{l} \rbrace}{(t-t')^{N^{U}_{2}}}e^{2\pi i \xi \sum_{k=1}^{2}\sum_{l=1}^{N^{U}_{2}}\tilde{g}^{k}_{\gamma^U_{l}}\tilde{g}^{k}_{\alpha}(\Theta(T^{U}_{l}-\frac{L}{\tilde{v}_{k}}+t)-\Theta(T^{U}_{l}-\frac{L}{\tilde{v}_{k}}+t'))}\nonumber\\
&=& \left[\int_{-t+\frac{L}{\tilde{v}_{2}}}^{-t'+\frac{L}{\tilde{v}_{2}}} \frac{d T^{U}_{l}}{t-t'} e^{2\pi i \xi \sum_{k=1}^{2} \tilde{g}^{k}_{1}\tilde{g}^{k}_{\alpha}(\Theta(T^{U}_{l}-\frac{L}{\tilde{v}_{k}}+t)-\Theta(T^{U}_{l}-\frac{L}{\tilde{v}_{k}}+t'))}\right]^{m^{U}_{2}}\nonumber\\
&& \times \left[\int_{-t+\frac{L}{\tilde{v}_{2}}}^{-t'+\frac{L}{\tilde{v}_{2}}} \frac{d T^{U}_{l}}{t-t'} e^{2\pi i \xi \sum_{k=1}^{2} \tilde{g}^{k}_{2}\tilde{g}^{k}_{\alpha}(\Theta(T^{U}_{l}-\frac{L}{\tilde{v}_{k}}+t)-\Theta(T^{U}_{l}-\frac{L}{\tilde{v}_{k}}+t'))}\right]^{N^{U}_{2}-m^{U}_{2}}\nonumber\\
I^L_{2} &=& \int_{-t+\frac{L}{\tilde{v}_{2}}}^{-t'+\frac{L}{\tilde{v}_{2}}} \frac{D\lbrace T^L_{l} \rbrace}{(t-t')^{N^L_{2}}}e^{-2\pi i \xi \sum_{k=1}^{2}\sum_{l=1}^{N^L_{2}}\tilde{g}^{k}_{\gamma^L_{l}}\tilde{g}^{k}_{\alpha}(\Theta(T^L_{l}-\frac{L}{\tilde{v}_{k}}+t)-\Theta(T^L_{l}-\frac{L}{\tilde{v}_{k}}+t'))}\nonumber\\
&=& \left[\int_{-t+\frac{L}{\tilde{v}_{2}}}^{-t'+\frac{L}{\tilde{v}_{2}}} \frac{d T^L_{l}}{t-t'} e^{-2\pi i \xi \sum_{k=1}^{2} \tilde{g}^{k}_{1}\tilde{g}^{k}_{\alpha}(\Theta(T^L_{l}-\frac{L}{\tilde{v}_{k}}+t)-\Theta(T^L_{l}-\frac{L}{\tilde{v}_{k}}+t'))}\right]^{m^L_{2}}\nonumber\\
&& \times \left[\int_{-t+\frac{L}{\tilde{v}_{2}}}^{-t'+\frac{L}{\tilde{v}_{2}}} \frac{d T^L_{l}}{t-t'} e^{-2\pi i \xi \sum_{k=1}^{2} \tilde{g}^{k}_{2}\tilde{g}^{k}_{\alpha}(\Theta(T^L_{l}-\frac{L}{\tilde{v}_{k}}+t)-\Theta(T^L_{l}-\frac{L}{\tilde{v}_{k}}+t'))}\right]^{N^L_{2}-m^L_{2}},    
\end{eqnarray}
with 
\begin{eqnarray}
f_{\alpha i}^{2 U/L}(\xi) &=& \int_{-t+\frac{L}{\tilde{v}_{2}}}^{-t'+\frac{L}{\tilde{v}_{2}}} \frac{d T^L_{l}}{t-t'} e^{\pm 2\pi i \xi \sum_{k=1}^{2} \tilde{g}^{k}_{i}\tilde{g}^{k}_{\alpha}(\Theta(T^L_{l}-\frac{L}{\tilde{v}_{k}}+t)-\Theta(T^L_{l}-\frac{L}{\tilde{v}_{k}}+t'))} = e^{\pm2\pi i\xi \Lambda^{2}_{\alpha i}}.
\end{eqnarray}

Now with this, the tunneling noise reads
\begin{eqnarray}
\langle S_{T}\rangle &=& \sum_{\alpha}q_{\alpha}^{2}|\Gamma_{\alpha}|^{2}\sum_{\xi}\sum_{N^{U/L}_{1}=0}^{\infty}\sum_{N^{U/L}_{2}=0}^{\infty}\sum_{m^{U/L}_{1}=0}^{N^{U/L}_{1}}\sum_{m^{U/L}_{2}=0}^{N^{U/L}_{2}} \frac{C^{N^{U}_{1}}_{m^{U}_{1}} C^{N^{U}_{2}}_{m^{U}_{2}} C^{N^L_{1}}_{m^L_{1}} C^{N^L_{2}}_{m^L_{2}}}{N^{U}_{1}! N^{U}_{2}! N^L_{1}! N^L_{2}!}\int_{-\infty}^{t} dt'  \nonumber\\
&& \times\left[ \prod_{k=1}^{2}(\tilde{G}^{-+}_{k}(t-t'))^{2(\tilde{g}^{k}_{\alpha})^{2}} + \prod_{k=1}^{2}(\tilde{G}^{-+}_{k}(t'-t))^{2(\tilde{g}^{k}_{\alpha})^{2}} \right]e^{-(\gamma^{U}_{1} + \gamma^L_{1} + \gamma^{U}_{2} + \gamma^L_{2})(t-t')} \nonumber\\
&& \times\left[\Theta(t-t'-\Delta\tau) (\gamma^{U}_{1} f_{\alpha 1}^{1U}(\xi)(t-t'))^{m^{U}_{1}}(\gamma^{U}_{2} f_{\alpha 2}^{1U}(\xi)(t-t'))^{N^{U}_{1}-m^{U}_{1}}(\gamma^{U}_{1}f_{\alpha 1}^{2U}(\xi)\Delta\tau)^{m^{U}_{2}}\right.\nonumber\\
&& \left.\times (\gamma^{U}_{2}f_{\alpha 2}^{2U}(\xi)\Delta\tau)^{N^{U}_{2}-m^{U}_{2}}(\gamma^L_{1}f_{\alpha 1}^{1D}(\xi)(t-t'))^{m^L_{1}}(\gamma^L_{2}f_{\alpha 2}^{1D}(\xi)(t-t'))^{N^L_{1}-m^L_{1}}(\gamma^L_{1}f_{\alpha 1}^{2D}(\xi)\Delta\tau)^{m^L_{2}}\right.\nonumber\\
&& \left. \times(\gamma^L_{2}f_{\alpha 2}^{2D}(\xi)\Delta\tau)^{N^L_{2}-m^L_{2}} e^{-(\gamma^{U}_{1} + \gamma^L_{1} + \gamma^{U}_{2} + \gamma^L_{2})(\Delta\tau)}\right.\nonumber\\
&& \left. + \Theta(\Delta\tau-(t-t'))  (\gamma^{U}_{1} f_{\alpha 1}^{1U}(\xi)(t-t'))^{m^{U}_{1}}(\gamma^{U}_{2} f_{\alpha 2}^{1U}(\xi)(t-t'))^{N^{U}_{1}-m^{U}_{1}}(\gamma^{U}_{1}f_{\alpha 1}^{2U}(\xi)(t-t')))^{m^{U}_{2}}\right.\nonumber\\
&& \left.\times  (\gamma^{U}_{2}f_{\alpha 2}^{2U}(\xi)(t-t'))^{N^{U}_{2}-m^{U}_{2}}(\gamma^L_{1}f_{\alpha 1}^{1D}(\xi)(t-t'))^{m^L_{1}}(\gamma^L_{2}f_{\alpha 2}^{1D}(\xi)(t-t'))^{N^L_{1}-m^L_{1}}(\gamma^L_{1}f_{\alpha 1}^{2D}(\xi)(t-t')))^{m^L_{2}}\right.\nonumber\\
&& \left.\times(\gamma^L_{2}f_{\alpha 2}^{2D}(\xi)(t-t')))^{N^L_{2}-m^L_{2}} e^{-(\gamma^{U}_{1} + \gamma^L_{1} + \gamma^{U}_{2} + \gamma^L_{2})(t-t')} \right].
\end{eqnarray}

\begin{eqnarray}
\langle S_{T}\rangle &=&  \sum_{\alpha}q_{\alpha}^{2}|\Gamma_{\alpha}|^{2}\sum_{\xi}\sum_{N^{U/L}_{1}=0}^{\infty}\sum_{N^{U/L}_{2}=0}^{\infty}\frac{1}{N^{U}_{1}! N^{U}_{2}! N^L_{1}! N^L_{2}!}\int_{-\infty}^{t} dt'  \nonumber\\
&& \times\left[ \prod_{k=1}^{2}(\tilde{G}^{-+}_{k}(t-t'))^{2(\tilde{g}^{k}_{\alpha})^{2}} + \prod_{k=1}^{2}(\tilde{G}^{-+}_{k}(t'-t))^{2(\tilde{g}^{k}_{\alpha})^{2}} \right]e^{-(\gamma^{U}_{1} + \gamma^L_{1} + \gamma^{U}_{2} + \gamma^L_{2})(t-t')} \nonumber\\
&& \times\left[ \Theta(t-t'-\Delta\tau) ((t-t')(\gamma^{U}_{1} f_{\alpha 1}^{1U}(\xi)+\gamma^{U}_{2} f_{\alpha 2}^{1U}(\xi)))^{N^{U}_1}(\Delta\tau(\gamma^{U}_{1} f_{\alpha 1}^{2U}(\xi)+\gamma^{U}_{2} f_{\alpha 2}^{2U}(\xi)))^{N^{U}_2} \right.\nonumber\\
&& \left. \times ((t-t')(\gamma^L_{1} f_{\alpha 1}^{1D}(\xi)+\gamma^L_{2} f_{\alpha 2}^{1D}(\xi)))^{N^L_1}(\Delta\tau(\gamma^L_{1} f_{\alpha 1}^{2D}(\xi)+\gamma^L_{2} f_{\alpha 2}^{2D}(\xi)))^{N^L_2}e^{-(\gamma^{U}_{1}+\gamma^{U}_{2}+\gamma^L_{1}+\gamma^L_{2})\Delta\tau}\right.\nonumber\\
&& \left. + \Theta(\Delta\tau-(t-t'))((t-t')(\gamma^{U}_{1} f_{\alpha 1}^{1U}(\xi)+\gamma^{U}_{2} f_{\alpha 2}^{1U}(\xi)))^{N^{U}_1}((t-t'))(\gamma^{U}_{1} f_{\alpha 1}^{2U}(\xi)+\gamma^{U}_{2} f_{\alpha 2}^{2U}(\xi)))^{N^{U}_2} \right.\nonumber\\
&& \left. \times ((t-t')(\gamma^L_{1} f_{\alpha 1}^{1D}(\xi)+\gamma^L_{2} f_{\alpha 2}^{1D}(\xi)))^{N^L_1}((t-t'))(\gamma^L_{1} f_{\alpha 1}^{2D}(\xi)+\gamma^L_{2} f_{\alpha 2}^{2D}(\xi)))^{N^L_2}e^{-(\gamma^{U}_{1}+\gamma^{U}_{2}+\gamma^L_{1}+\gamma^L_{2})(t-t'))}\right]\nonumber\\
&=& \sum_{\alpha}q_{\alpha}^{2}|\Gamma_{\alpha}|^{2}\sum_{\xi}\int_{-\infty}^{t} dt' \left[ \prod_{k=1}^{2}(\tilde{G}^{-+}_{k}(t-t'))^{2(\tilde{g}^{k}_{\alpha})^{2}} + \prod_{k=1}^{2}(\tilde{G}^{-+}_{k}(t'-t))^{2(\tilde{g}^{k}_{\alpha})^{2}} \right] \nonumber\\
&& \times e^{-(\gamma^{U}_{1} + \gamma^L_{1} + \gamma^{U}_{2} + \gamma^L_{2})(t-t')} \nonumber\\
&& \times\left[ \Theta(t-t'-\Delta\tau) \mathrm{Exp}\left[ (t-t')(\gamma^{U}_{1} f_{\alpha 1}^{1U}(\xi)+\gamma^{U}_{2} f_{\alpha 2}^{1U}(\xi) + \gamma^L_{1} f_{\alpha 1}^{1D}(\xi)+\gamma^L_{2} f_{\alpha 2}^{1D}(\xi))\right.\right.\nonumber\\
&& \left.\left.+ \Delta\tau (\gamma^{U}_{1} f_{\alpha 1}^{2U}(\xi)+\gamma^{U}_{2} f_{\alpha 2}^{2U}(\xi) + \gamma^L_{1} f_{\alpha 1}^{2D}(\xi)+\gamma^L_{2} f_{\alpha 2}^{2D}(\xi)) \right]e^{-(\gamma^{U}_{1}+\gamma^{U}_{2}+\gamma^L_{1}+\gamma^L_{2})\Delta\tau}\right.\nonumber\\
&& \left. + \Theta(\Delta\tau-(t-t'))\mathrm{Exp}\left[ (t-t')(\gamma^{U}_{1} f_{\alpha 1}^{1U}(\xi)+\gamma^{U}_{2} f_{\alpha 2}^{1U}(\xi) + \gamma^L_{1} f_{\alpha 1}^{1D}(\xi)+\gamma^L_{2} f_{\alpha 2}^{1D}(\xi)\right.\right.\nonumber\\
&& \left.\left.+ \gamma^{U}_{1} f_{\alpha 1}^{2U}(\xi)+\gamma^{U}_{2} f_{\alpha 2}^{2U}(\xi) + \gamma^L_{1} f_{\alpha 1}^{2D}(\xi)+\gamma^L_{2} f_{\alpha 2}^{2D}(\xi)) \right]e^{-(\gamma^{U}_{1}+\gamma^{U}_{2}+\gamma^L_{1}+\gamma^L_{2})(t-t')}\right]
\end{eqnarray}
Now keeping in mind the fact that $f_{\alpha i}^{jU}(\xi) = (f_{\alpha i}^{jD}(\xi))^{*}$, we can write the above equation as
\begin{eqnarray}
\langle S_{T}\rangle &=& 2\sum_{\alpha}q_{\alpha}^{2}|\Gamma_{\alpha}|^{2}\int_{0}^{\infty} dt \left[ \prod_{k=1}^{2}(\tilde{G}^{-+}_{k}(t))^{2(\tilde{g}^{k}_{\alpha})^{2}} + \prod_{k=1}^{2}(\tilde{G}^{-+}_{k}(-t))^{2(\tilde{g}^{k}_{\alpha})^{2}} \right]  \nonumber\\
&& \times \left[\frac{\Theta(t-\Delta\tau)\cos[\gamma^{-}_{1}(t \mathrm{Im}(f^{1}_{\alpha 1}) + \Delta\tau \mathrm{Im}(f^{2}_{\alpha 1})) + \gamma^{-}_{2}(t \mathrm{Im}(f^{1}_{\alpha 2}) + \Delta\tau \mathrm{Im}(f^{2}_{\alpha 2}))]}{\mathrm{Exp}\left[ \gamma^{+}_{1}(t(1-\mathrm{Re}(f^{1}_{\alpha 1})) + \Delta\tau(1-\mathrm{Re}(f^{2}_{\alpha 1}) )) +  \gamma^{+}_{2}(t(1-\mathrm{Re}(f^{1}_{\alpha 2})) + \Delta\tau(1-\mathrm{Re}(f^{2}_{\alpha 2})) )\right]}\right.\nonumber\\
&& \left. + \frac{\Theta(\Delta\tau-t)\cos[t(\gamma^{-}_{1}( \mathrm{Im}(f^{1}_{\alpha 1}) + \mathrm{Im}(f^{2}_{\alpha 1})) + \gamma^{-}_{2}( \mathrm{Im}(f^{1}_{\alpha 2}) +  \mathrm{Im}(f^{2}_{\alpha 2}))]}{\mathrm{Exp}\left[ t(\gamma^{+}_{1}(2-\mathrm{Re}(f^{1}_{\alpha 1})-\mathrm{Re}(f^{2}_{\alpha 1}) ) +  \gamma^{+}_{2}(2-\mathrm{Re}(f^{1}_{\alpha 2}) -\mathrm{Re}(f^{2}_{\alpha 2})) )\right]} \right]
\label{Eq:Poissonian_averaged_noise}
\end{eqnarray}

In order to arrive at Eq.~\eqref{Eq:Poissonian_averaged_noise}, we made the transformation $t-t'\longrightarrow t$ and taken $\gamma^{\pm}_{i} = \gamma^{U}_{i} \pm \gamma^L_{i}$. Also note that $f^{1U}_{\alpha i}(\xi) = f^{1D}_{\alpha i}(-\xi) = f^{1}_{\alpha i}$ and $f^{2U}_{\alpha i}(\xi) = f^{2D}_{\alpha i}(-\xi) = f^{2}_{\alpha i}$.

A very similar calculation can be attempted to calculate the Poissonian averaged tunneling current, which results in
\begin{eqnarray}
\langle I_{T}(t) \rangle &=& 2i\sum_{\alpha}q_{\alpha}|\Gamma_{\alpha}|^{2}\int_{0}^{\infty} dt \left[ \prod_{k=1}^{2}(\tilde{G}^{-+}_{k}(t))^{2(\tilde{g}^{k}_{\alpha})^{2}} - \prod_{k=1}^{2}(\tilde{G}^{-+}_{k}(-t))^{2(\tilde{g}^{k}_{\alpha})^{2}} \right]  \nonumber\\
&& \times\left[\frac{\Theta(t-\Delta\tau)\sin[\gamma^{-}_{1}(t \mathrm{Im}(f^{1}_{\alpha 1}) + \Delta\tau \mathrm{Im}(f^{2}_{\alpha 1})) + \gamma^{-}_{2}(t \mathrm{Im}(f^{1}_{\alpha 2}) + \Delta\tau \mathrm{Im}(f^{2}_{\alpha 2}))]}{\mathrm{Exp}\left[ \gamma^{+}_{1}(t(1-\mathrm{Re}(f^{1}_{\alpha 1})) + \Delta\tau(1-\mathrm{Re}(f^{2}_{\alpha 1}) )) +  \gamma^{+}_{2}(t(1-\mathrm{Re}(f^{1}_{\alpha 2})) + \Delta\tau(1-\mathrm{Re}(f^{2}_{\alpha 2})) )\right]}\right.\nonumber\\
&& \left.+\frac{\Theta(\Delta\tau-t)\sin[t(\gamma^{-}_{1}( \mathrm{Im}(f^{1}_{\alpha 1}) + \mathrm{Im}(f^{2}_{\alpha 1})) + \gamma^{-}_{2}( \mathrm{Im}(f^{1}_{\alpha 2}) +  \mathrm{Im}(f^{2}_{\alpha 2}))]}{\mathrm{Exp}\left[ t( \gamma^{+}_{1}(2-\mathrm{Re}(f^{1}_{\alpha 1})-\mathrm{Re}(f^{2}_{\alpha 1}) ) +  \gamma^{+}_{2}(2-\mathrm{Re}(f^{1}_{\alpha 2})-\mathrm{Re}(f^{2}_{\alpha 2})) )\right]}\right]
\label{Eq:Poissonian_averaged_tunneling_current}
\end{eqnarray}

The average injected current across the source QPC can be expressed as $I^{U/L}=I^{U/L}_{1} + I^{U/L}_{2}$, where $I^{U/L}_{i}$ is the average injected current across the source QPCs corresponding to the $i^{th}$-type anyon in the upper/lower edge. Since the charge of both types of excitation is the same ($e^* = e/5$), the injected current $I^{U/L}_{i}$ can be written as $I^{U/L}_{i} = P_{i}I^{U/L}$, where $P_{i}$ is the proportion of the $i^{th}$-type anyons among all the injected anyons in the upper/lower edge, such that $\sum_{i}P_{i}=1$. Here, the $P_{i}$ is taken to be the same for both the source QPCs. The injection rates are then given by $\gamma^{U/L}_{i} = I^{U/L} P_{i}/q$ (as the charge $q=1/5$ is the same for both types of anyon). With this, $\gamma^{\pm}_{i} = I^{\pm}P_{i}/q$, where $I^{\pm} = I^{U}\pm I^L$. Now we can express the Poissonian averaged tunneling current and tunneling noise as  
\begin{eqnarray}
\langle I_{T}(t) \rangle &=& 2i\sum_{\alpha}q_{\alpha}|\Gamma_{\alpha}|^{2}\int_{0}^{\infty} dt \left[ \prod_{k=1}^{2}(\tilde{G}^{-+}_{k}(t))^{2(\tilde{g}^{k}_{\alpha})^{2}} - \prod_{k=1}^{2}(\tilde{G}^{-+}_{k}(-t))^{2(\tilde{g}^{k}_{\alpha})^{2}} \right]  \nonumber\\
&& \times\left(\frac{\Theta(\Delta\tau-t)\sin[\frac{I_{-}}{q}t\sum_{\gamma=1}^{2}P_{\gamma}(\sin(2\pi \Lambda^{1}_{\alpha \gamma}) + \sin(2\pi \Lambda^{2}_{\alpha \gamma}))]}{\mathrm{Exp}\left[\frac{I_{+}}{q}t\sum_{\gamma=1}^{2} P_{\gamma}(2 - (\cos[2\pi\Lambda^{1}_{\alpha \gamma}] + \cos[2\pi\Lambda^{2}_{\alpha \gamma}]))\right]}\right.\nonumber\\
&& \left. + \frac{\Theta(t-\Delta\tau)\sin[\frac{I_{-}}{q}\sum_{\gamma=1}^{2}P_{\gamma}(\Delta\tau(\sin(2\pi \Lambda^{1}_{\alpha \gamma}) + \sin(2\pi \Lambda^{2}_{\alpha \gamma})) + (t-\Delta\tau)\sin(2\pi \textbf{g}_{\alpha}.\textbf{g}_{\gamma}))]}{\mathrm{Exp}\left[\frac{I_{+}}{q}\sum_{\gamma=1}^{2} P_{\gamma}(t+\Delta\tau - \Delta\tau(\cos[2\pi\Lambda^{1}_{\alpha \gamma}] + \cos[2\pi\Lambda^{2}_{\alpha \gamma}]) -(t-\Delta\tau)\cos[2\pi\textbf{g}_{\alpha}.\textbf{g}_{\gamma}]) \right]}\right) ,\\
\langle S_{T}(t) \rangle &=& 2\sum_{\alpha}q^{2}_{\alpha}|\Gamma_{\alpha}|^{2}\int_{0}^{\infty} dt \left[ \prod_{k=1}^{2}(\tilde{G}^{-+}_{k}(t))^{2(\tilde{g}^{k}_{\alpha})^{2}} + \prod_{k=1}^{2}(\tilde{G}^{-+}_{k}(-t))^{2(\tilde{g}^{k}_{\alpha})^{2}} \right]  \nonumber\\
&& \times\left(\frac{\Theta(\Delta\tau-t)\cos[\frac{I_{-}}{q}t\sum_{\gamma=1}^{2}P_{\gamma}(\sin(2\pi \Lambda^{1}_{\alpha \gamma}) + \sin(2\pi \Lambda^{2}_{\alpha \gamma}))]}{\mathrm{Exp}\left[\frac{I^{+}}{q}t\sum_{\gamma=1}^{2} P_{\gamma}(2 - (\cos[2\pi\Lambda^{1}_{\alpha \gamma}] + \cos[2\pi\Lambda^{2}_{\alpha \gamma}]))\right]}\right.\nonumber\\
&& \left. + \frac{\Theta(t-\Delta\tau)\cos[\frac{I_{-}}{q}\sum_{\gamma=1}^{2}P_{\gamma}(\Delta\tau(\sin(2\pi \Lambda^{1}_{\alpha \gamma}) + \sin(2\pi \Lambda^{2}_{\alpha \gamma})) + (t-\Delta\tau)\sin(2\pi \textbf{g}_{\alpha}.\textbf{g}_{\gamma}))]}{\mathrm{Exp}\left[\frac{I_{+}}{q}\sum_{\gamma=1}^{2} P_{\gamma}(t+\Delta\tau - \Delta\tau(\cos[2\pi\Lambda^{1}_{\alpha \gamma}] + \cos[2\pi\Lambda^{2}_{\alpha \gamma}]) -(t-\Delta\tau)\cos[2\pi\textbf{g}_{\alpha}.\textbf{g}_{\gamma}]) \right]}\right).
\label{Eq:Final_T_current_noise1}
\end{eqnarray}
Here, $\Lambda^{k}_{\alpha \gamma} = \tilde{g}^{k}_{\alpha}\tilde{g}^{k}_{ \gamma}$. Note that $\Lambda^{1}_{\alpha  \gamma}+\Lambda^{2}_{\alpha  \gamma} = \textbf{g}_{\alpha}.\textbf{g}_{ \gamma}$ is an invariant braiding phase which is independent of the interaction strength. Equation~\eqref{Eq:Final_T_current_noise1} is very intuitive. 
Due to the inter-edge interaction, we have different renormalized velocities corresponding to the two edge modes. This results in a time lag of $\Delta\tau$, which affects the expressions for tunneling noise and tunneling current. The integrand in Eq.\eqref{Eq:Final_T_current_noise1} is divided into two time domains: $t<\Delta\tau$ and $t>\Delta\tau$. For the time period $t<\Delta\tau$, the contribution comes only from the uncorrelated part of the Poissonian distribution, leading to the interaction-dependent braiding phase of $\sin(2\pi\Lambda^{1}_{\alpha \gamma}) + \sin(2\pi\Lambda^{2}_{\alpha \gamma})$. However, for the time domain of $t>\Delta\tau$, the contribution to the braiding phase comes from both the two uncorrelated parts, as well as from the correlated part of the Poissonian stream of anyons. This leads to an interaction-dependent braiding phase of $\Delta\tau(\sin(2\pi\Lambda^{1}_{\alpha \gamma}) + \sin(2\pi\Lambda^{2}_{\alpha \gamma})) + (t-\Delta\tau)\sin(2\pi \textbf{g}_{\alpha}.\textbf{g}_{\gamma})$.